\documentclass{article}

\usepackage{jheppub}

\usepackage{tikz}

\newcommand\de{\text{d}}

\title{Interpolating families of integrable AdS$_{\mathbf{3}}$ backgrounds}

\author[a]{Fiona~K.~Seibold,}
\author[b]{Alessandro Sfondrini%
}

\affiliation[a]{Deutsches Elektronen-Synchrotron DESY, Notkestr. 85, 22607 Hamburg, Germany}
\affiliation[b]{Dipartimento di Fisica e Astronomia, Universit\`a degli Studi di Padova,\\
\& Istituto Nazionale di Fisica Nucleare, Sezione di Padova,\\
via Marzolo 8, 35131 Padova, Italy}

\emailAdd{fiona.seibold@desy.de}
\emailAdd{alessandro.sfondrini@unipd.it}

\abstract{
We construct families of integrable deformations that interpolate between $AdS_3\times S^3\times S^3\times S^1$ and either $AdS_3\times S^3\times S^2\times T^2$ or $AdS_3\times S^2\times S^2\times T^3$. They preserve half of the supersymmetry of the original background, namely one copy of the $\mathfrak{d}(2,1;\alpha)$ algebra. From this it follows a similar integrable interpolation between $AdS_3\times S^3\times T^4$ and $AdS_3\times S^2\times T^5$, which also preserves half of the supersymmetry, namely a copy of the $\mathfrak{psu}(1,1|2)$ algebra. In all cases, the interpolating backgrounds are constructed by using TsT transformations, which makes it easy to implement them in the integrability formalism in the full quantum theory. To illustrate this point, we discuss the lightcone gauge fixing of the models and compute their pp-wave Hamiltonian.
}

\begin{document} 
\begin{flushright}
    DESY-25-036
\end{flushright}
\maketitle
\flushbottom
\newpage


\section{Introduction}
Integrability is a powerful tool to study the classical and quantum dynamics of strings on certain curved backgrounds, especially in the context of the AdS/CFT correspondence~\cite{Maldacena:1997re}. Integrability appears in \textit{perturbative} string theory, when the string interaction coupling is vanishingly small, but the string tension is arbitrary --- this corresponds to the planar / large-$N$ limit in the AdS/CFT correspondence. 
After yielding remarkable results in the context of the maximally supersymmetric $AdS_5$/CFT$_4$ duality~\cite{Arutyunov:2009ga,Beisert:2010jr}, integrability techniques have been applied to other setups. In particular, the $AdS_3$/CFT$_2$ correspondence proved to be a challenging and very rewarding landscape: it has half the amount of supersymmetry (16 Killing spinors, instead of the 32 Killing spinors of $AdS_5\times S^5$) and therefore more moduli than $AdS_5\times S^5$. Importantly, $AdS_3$ superstring backgrounds can be supported by a mixture of Ramond-Ramond (RR) and Neveu-Schwarz-Neveu-Schwarz (NSNS) fluxes, and remain integrable throughout~\cite{Cagnazzo:2012se}, see~\cite{Sfondrini:2014via, Demulder:2023bux} for reviews. The maximally supersymmetric $AdS_3$ backgrounds fall in three families: $AdS_3\times S^3\times T^4$, $AdS_3\times S^3\times K3$ and $AdS_3\times S^3\times S^3\times S^1$. The first two families are more closely related: they both have only one curvature scale (the supergravity equations require the radius of $AdS_3$ to be the same as that of $S^3$), and in both cases superisometries include the $\mathfrak{psu}(1,1|2)_L\oplus \mathfrak{psu}(1,1|2)_R$ superalgebra --- $L,R$ denote the left and right copy of the isometries, corresponding to the chiral split of the supersymmetries as $\mathcal{N}=(4,4)$ in the dual superconformal field theory.  There are still important differences starting, for instance, from the BPS spectrum of the models which crucially depends on the choice of the hyperk\"ahler manifold. In the context of integrability, most of the efforts have so far been devoted to understanding the~$T^4$ case~\cite{Ekhammar:2021pys,Cavaglia:2021eqr,Frolov:2021bwp,Brollo:2023pkl}.%
\footnote{Actually, most of the literature focuses on a sector of the string spectrum that is insensitive to the global boundary conditions of the flat geometry, effectively treating $T^4$ as if it were~$\mathbb{R}^4$; see~\cite{OhlssonSax:2018hgc} for a discussion how which moduli turn out to be important in this sector.}
The case of  $AdS_3\times S^3\times S^3\times S^1$, whose integrability was first discussed in~\cite{Babichenko:2009dk}, is in a sense the richest of the three: it depends on an additional parameter --- the relative size of the two~$S^3$ --- and correspondingly its supersymmetries  are given by $\mathfrak{d}(2,1;\alpha)_L\oplus\mathfrak{d}(2,1;\alpha)_R$,%
\footnote{The exceptional Lie superalgebra $\mathfrak{d}(2,1;\alpha)$ has R-symmetry $\mathfrak{su}(2)\oplus\mathfrak{su}(2)$, and in the limit $\alpha\to0$ or $\alpha\to1$ yields $\mathfrak{psu}(1,1|2)$ as a contraction. The dual $\mathcal{N}=(4,4)$ superconformal symmetry for $\mathfrak{d}(2,1;\alpha)$ is called ``large'' to distinguish it from the ``small'' superconformal symmetry related to $\mathfrak{psu}(1,1|2)$.} 
where $0<\alpha<1$ is a free parameter. More specifically, calling $R$ the radius of $AdS_3$ and $R_{1},R_{2}$ the radius of either three-sphere, we have from the supergravity equations
\begin{equation}
    \frac{1}{R^2}=\frac{1}{R_1^2}+\frac{1}{R_2^2}\,,\qquad
    R_1 = \frac{R}{\sqrt{\alpha}}\,, \quad R_2 = \frac{R}{\sqrt{1-\alpha}}\,,\qquad
    0<\alpha<1\,.
\end{equation}
Therefore, if we ignore the global topology of the hyperk\"aler manifold, we can think that taking $\alpha\to0$ or $\alpha\to1$ yields back $AdS_3\times S^3\times T^4$.
Hence, not only strings on $AdS_3\times S^3\times S^3\times S^1$ are richer than their $T^4$ counterpart, but many lessons learned from them might straightforwardly apply to the  $T^4$  case too by taking the limit $\alpha\to0$ (or $\alpha\to1$). It is in this sprit that here we study the deformations of $AdS_3\times S^3\times S^3\times S^1$ backgrounds.

In general, integrable deformations of a supersymmetric background break most of its supersymmetry, see~\cite{Hoare:2021dix,Seibold:2024qkh} for reviews.
In the case of $AdS_3$ backgrounds however, because the supersymmetry algebra has a factorised form, it is possible to perform a wealth of deformations without affecting one half of the supersymmetries. Using this logic, supersymmetric integrable deformations of the $AdS_3 \times S^3 \times T^4$ superstring were constructed in~\cite{Hoare:2022asa}, preserving 8~Killing spinors.%
\footnote{Before that, the supersymmetry-preserving deformations of $AdS_3\times S^3\times T^4$ had been constructed starting from the brane realisation of the model; they turned out to be equivalent to TsT transformations and to preserve half of the supersymmetry~\cite{Orlando:2010ay,Orlando:2010yh,Orlando:2012hu}; see also~\cite{Israel:2004vv} for related earlier work in the worldsheet-CFT approach.}
In particular, the authors considered the effect of certain TsT transformations~\cite{Lunin:2005jy}. This is a sequence of transformations which requires picking two $\mathfrak{u}(1)$ isometries, call them $A$ and~$B$: first, a T-duality along the $\mathfrak{u}(1)_A$ direction, then shift along $\mathfrak{u}(1)_B$ direction, finally a T-duality ``back'' along  $\mathfrak{u}(1)_A$.
This sequence of operations yields a new background which, if $\mathfrak{u}(1)_A$ or $\mathfrak{u}(1)_B$ are Cartan generators of a nonabelian isometry, has fewer symmetries than the original one. Still, the transformation preserves integrability and acts on the worldsheet S~matrix as a Drinfel'd-Reshetikin twist~\cite{Frolov:2005dj,Alday:2005ww} (see also~\cite{vanTongeren:2021jhh} for a detailed discussion of the twist).
In~\cite{Hoare:2022asa}, the authors considered TsT transformations constructed out of Cartan generators sitting in a copy of $\mathfrak{psu}(1,1|2)$, thereby preserving the other copy of $\mathfrak{psu}(1,1|2)$.
Interestingly, they found that the resulting geometry is regular, has a constant dilaton, and it interpolates between  $AdS_3\times S^3\times T^4$ and $AdS_2\times S^2\times T^6$ if one performs an analytic continuation (and if we are cavalier in treating $\mathbb{R}^n$ and $T^n$ on the same footing). For generic values of the deformation parameter, the geometry involves a squashed sphere and a warped $AdS_3$ space. 
The purpose of this paper is to investigate whether similar deformations exist which interpolates between an $S^3$ and an $S^2\times \mathbb{R}$ internal space while preserving eight Killing spinors \textit{and without spoiling the $AdS_3$ factor}.%
\footnote{
Deformations preserving the $AdS_d$ part of the geometry were considered in the context of $\lambda$-deformations in~\cite{Itsios:2019izt,Itsios:2024ybt}.
}
The reason for the latter requirement is that, hopefully, the interpretation of the deformation should be clearer in in the context of the dual CFT$_2$. 
At this point let us mention that the pure-NSNS $AdS_3 \times S^3 \times T^4$ and $AdS_3 \times S^3 \times S^3 \times S^1$ theories have a simple description in the RNS formalism as Wess-Zumino-Witten models~\cite{Maldacena:2000hw} (as well as a simple integrability description~\cite{Baggio:2018gct,Dei:2018mfl,Dei:2018jyj}). Various deformations of the pure-NSNS theories were constructed and studied in~\cite{Borsato:2018spz,Eloy:2024lwn}. These deformations remain pure-NSNS, include TsT transformations and may be supersymmetric. In view of our motivation to interpolate between known integrable backgrounds which generically have RR fluxes, we will consider generic mixed-flux theories; as we will see, the presence of RR flux is important to yield an interpolation to a different geometry.
It is natural to start from the $AdS_3 \times S^3 \times S^3 \times S^1$ geometry (with RR and NSNS flux). Then, we have several options for TsT transformations which preserve the $AdS_3$ factor as well as half of the supersymmetry --- let us say, preserve $\mathfrak{d}(2,1;\alpha)_L$. We can use the Cartan elements belonging to the two $\mathfrak{su}(2)$ algebras within $\mathfrak{d}(2,1;\alpha)_R$, or the $\mathfrak{u}(1)$ relative to the $S^1$ factor. It turns out that the most general TsT transformation has three parameters, on top of the original parameters of the  $AdS_3 \times S^3 \times S^3 \times S^1$ background, which are the $AdS_3$ radius (in units of the string length), the parameter $\alpha$, and the relative strength of RR vs.~NSNS fluxes. This gives a very rich scenario, which we will explore in this paper by constructing the relevant superstring backgrounds (including all fluxes) and studying them in the pp-wave limit~\cite{Blau:2002dy,Berenstein:2002jq}.

The plan of this paper is as follows. We start in Section~\ref{sec:overview} by giving a brief overview of the known integrable string backgrounds (found elsewhere~\cite{Wulff:2017zbl,Wulff:2017vhv}) among which our deformations will interpolate. Then, we study three different cases: the interpolation from $AdS_3 \times S^3 \times S^3 \times S^1$ to $AdS_3 \times S^3 \times S^2 \times T^2$ in Section~\ref{sec:AB}, the interpolation to  $AdS_3 \times S^2 \times S^2 \times T^3$ in Section~\ref{sec:AC}, and finally the most general three-parameter deformation in Section~\ref{sec:ABC}. While the last case does include the previous two by taking suitable values of the parameters, it is quite convoluted and we find it advantageous to discuss the important physical points of our constructions on the more transparent first two cases; indeed, the last case is included mostly for the sake of completeness. For each of these three cases we discuss the geometry emerging from the TsT transformation, the choice of a suitable (1/8-BPS) geodesics for the pp-wave expansion, and then we compute the pp-wave geometry and Hamiltonian, including the dispersion relation. We briefly comment on the analogy between these backgrounds and those emerging from trigonometric deformations in Section~\ref{sec:trigonometric}.
We conclude in Section~\ref{sec:conclusions}. Some more cumbersome expressions, related to Section~\ref{sec:ABC}, can be found in appendix~\ref{app:3param}.

\begin{table}[t]
\begin{center}
\renewcommand{\arraystretch}{1.5}
\begin{tabular}{c|cccc}
     Case & Geometry & SUSY & Manifest symmetries & Parameters \\ \hline
    1. & $AdS_3 \times S^3 \times S^3 \times S^1$ & 16 & $D(2,1;\alpha)^2 \times U(1)$ & 2 \\
    2. & $AdS_3 \times S^3 \times S^2 \times T^2$ & 8 & $D(2,1;\alpha) \times SL(2;\mathbb{R}) \times SU(2) \times U(1)^2$ & 1 \\
    3. & $AdS_3 \times S^2 \times S^2 \times T^3$ & 8 & $D(2,1;\alpha) \times SL(2;\mathbb{R})  \times U(1)^3$ & 1 \\ \hline
    4. & $AdS_3 \times S^3 \times T^4$ & 16 & $PSU(1,1|2)^2 \times U(1)^4$ & 2  \\
    5. & $AdS_3 \times S^2 \times T^5$ & 8 & $PSU(1,1|2) \times SU(2) \times U(1)^5$ & 0  \\ \hline
\end{tabular}
\end{center}
    \caption{All integrable symmetric-space type-IIB backgrounds with an $AdS_3$ factor. We also list the amount of SUSY (number of Killing spinors), the super-isometries, and how many free parameters there are in the fluxes aside from the overall scale --- i.e., the radius~$R$ of $AdS_3$, see e.g.~\eqref{eq:generalmetric}.}
    \label{tab:backgrounds}
\end{table}

\section{Integrable \texorpdfstring{AdS$_{\mathbf{3}}$}{AdS3} string backgrounds}
\label{sec:overview}

All symmetric-space solutions of type II supergravity with an  $AdS_3$ factor have been constructed in \cite{Wulff:2017zbl}, and their integrability was studied in~\cite{Wulff:2017vhv}.
The supersymmetric solutions take the schematic form
\begin{equation}
    AdS_{3}\times S^{3-n'}\times S^{3-n''}\times T^{n}\,,\qquad
    n-n'-n''=1\,,\quad n=1,\dots,5,
\end{equation}
where $n'$, $n''$ are non-negative integers. We label these five cases by the integer $n$, and summarise the backgrounds in Table~\ref{tab:backgrounds}.
The metric for the backgrounds 1., 2., 3.\ can be written in the form
\begin{equation}
\label{eq:generalmetric}
    ds^2 = R^2 \de s^2(AdS_3) + R_1^2 \de s^2 (S_1^{3-n'}) + R_2^2 \de s^2(S_2^{3-n''}) + \de s^2(T^{n})~,
\end{equation}
where $n'$ and $n''$ take values $0$ or~$1$, $R$ is the radius of $AdS_3$, $R_1$ and $R_2$ are the radii of the two three-spheres. To have a geometry with vanishing curvature, the radii obey the relations
\begin{equation}
\label{eq:metric123}
\begin{aligned}
    &1.: &\qquad R_1 &= \frac{R}{\sqrt{\alpha}}~, &\qquad R_2 &= \frac{R}{\sqrt{1-\alpha}}\,, \\
    &2.:  &\qquad R_1 &= \frac{R}{\sqrt{\alpha}}~, &\qquad R_2 &= \frac{R}{2 \sqrt{1-\alpha}}\,, \\
    &3.: &\qquad  R_1 &= \frac{R}{2 \sqrt{\alpha}}~, &\qquad R_2 &= \frac{R}{2 \sqrt{1-\alpha}}\,. 
    \end{aligned}
\end{equation}
The parameter~$\alpha$ with $0<\alpha<1$ is the same which appears in the exceptional Lie superalgebra $\mathfrak{d}(2,1;\alpha)$ describing the isometries of the space, cf.\ Table~\ref{tab:backgrounds}. In the limits $\alpha\to0$ and $\alpha\to1$ either $S^3$ (respectively, $S^2$) decompactifies to $\mathbb{R}^3$ (respectively, $\mathbb{R}^2$). 
In type IIB supergravity, the metric is supported by a Neveu-Schwarz-Neveu-Schwarz (NSNS) three-form flux $H_3$ related to the B-field as $H_3 = \de B_2$, as well as Ramond-Ramond (RR) fluxes $F_1$, $F_3$ and a self-dual five-form flux $F_5$. For the background 1., the fluxes are given by (for simplicity we set the dilaton $\Phi=0$)
\begin{equation}
\label{eq:background1}
    1.: \qquad  \left\{
    \begin{aligned}
        H_3 &= 2 q \left(R^2\, \Omega(AdS_3) + R_1^2\, \Omega(S_1^3) + R_2^2\,\Omega(S_2^3)\right) \\
        F_3 &= 2 \hat{q} \left(R^2\, \Omega(AdS_3) + R_1^2\, \Omega(S_1^3) + R_2^2\, \Omega(S_2^3)\right) \\
    \end{aligned} \right.
\end{equation}
where $\Omega(\mathcal X)$ denotes the unit-normalised volume form on the manifold $\mathcal X$, and $q$, $\hat{q}$ are parameters satisfying
\begin{equation}
    0\leq q\leq 1\,,\quad
    0\leq \hat{q}\leq 1\,,\qquad
    q^2 + \hat{q}^2=1\,.
\end{equation}
The appearance of the additional parameter $q$ indicates how the metric can be supported by a mixture of NSNS and RR fluxes: the point $q=1$ corresponds to the pure NSNS theory, while $q=0$ is the pure RR theory. The backgrounds 2. and 3. also feature a mixture of NSNS and RR fluxes, but their relative strength is fixed, and no additional parameter appears. We have
\begin{equation}
\label{eq:background2}
    2.: \qquad \left\{\begin{aligned}
        H_3 &= R_2 \Omega(S_2^2) \wedge \de x_1  \\
        F_3 &= R_2 \Omega(S_2^2) \wedge \de x_2 \\
        F_5 &= 2 \frac{R_1^3 R_2^2}{R}(1+\star) \Omega(S_1^3) \wedge \Omega(S_2^2) + 2 R_1^2 (1+\star) \Omega(S_1^3) \wedge \de x_1 \wedge \de x_2
    \end{aligned}\right.
\end{equation}
and
\begin{equation}
\label{eq:background3}
    3.: \qquad \left\{\begin{aligned}
        H_3 &= R_1 \Omega(S_1^2) \wedge \de x_2 + R_2 \Omega(S_2^2) \wedge \de x_3  \\
        F_3 &= R_1 \Omega(S_1^2) \wedge \de x_3 - R_2 \Omega(S_2^2) \wedge \de x_2 \\
        F_5 &= 2 \frac{R_1 R_2}{R} (1+\star) \Omega(S_1^2) \wedge \Omega(S_2^2) \wedge \de x_1
    \end{aligned}\right.
\end{equation}
The backgrounds 4.\ and 5.\ can be obtained from the cases 1., 2., 3.\ through the contraction limit $\alpha \rightarrow 1$ (or $\alpha \rightarrow 0$).

In this paper we will construct integrable backgrounds that interpolate between these geometries. We hence focus on the interpolating families which preserve the $AdS_3$ factor, and as much supersymmetry as possible (that is, eight supercharges).
To fix our notation, let us start from case 1., from which everything else can be in principle obtained (as we shall see).
Both $AdS_3$ and $S^3$ are group manifolds which therefore admit a left and right action of the (super)isometries. In particular, the $AdS_3$ factor has bosonic isometries given by $SU(1,1)_L \times SU(1,1)_R$ and each sphere has $SU(2)_{j,L} \times SU(2)_{j,R}$ where $j=1,2$ distinguish which sphere we are considering. All in all, we label the bosonic symmetries of $AdS_3 \times S^3 \times S^3 \times S^1$ as
\begin{equation}
     SU(1,1)_L \times SU(1,1)_R \times SU(2)_{1,L} \times SU(2)_{1,R} \times SU(2)_{2,L} \times SU(2)_{2,R} \times U(1)\,.
\end{equation}
Taking into account the Killing spinors, the $AdS_3 \times S^3 \times S^3 \times S^1$ (super)isometries combine into the supergroup
\begin{equation}
    D(2,1;\alpha)_L \times D(2,1;\alpha)_R \times U(1)\,,
\end{equation}
where the parameter $\alpha$ was introduced in eq.~\eqref{eq:metric123}.

\section{Integrable deformations from TsT transformations}
An important class of integrability-preserving transformations is given by the TsT family~\cite{Lunin:2005jy,Frolov:2005dj,Alday:2005ww}. A TsT transformation~\cite{Lunin:2005jy} is a combination of T-dualities and shifts which can be implemented for any string theory whose background presents at least two isometries, realised as shifts in the coordinates $x$ and $y$. The TsT transformation consists in performing a T-duality $x \rightarrow \tilde{x}$ (here the tilde denotes the T-dual coordinate), followed by a shift $y \rightarrow x + s \tilde{x}$, and finally another T-duality $\tilde{x} \rightarrow \tilde{\tilde{x}}$. A T-duality is an exact symmetry of string theory, and maps type IIB solutions to type IIA, and vice-versa. Therefore, since TsT transformations involve two T-dualities, they map e.g.\ type IIB solutions to type IIB solutions. Moreover, this type of deformation preserves the integrability of the model, and they can be interpreted as a state-dependent twist of the boundary conditions for the fields in the model~\cite{Frolov:2005dj,Alday:2005ww}.

Let us consider the $AdS_3 \times S_1^3 \times S_2^3 \times S^1$ metric in the Hopf parametrisation. Having stripped out the radii like in~\eqref{eq:generalmetric}, we have for each factor
\begin{align}
\de s^2(AdS_3) &= \frac{1}{4} \left( \de \vartheta^2 + \sinh^2 \vartheta \de \chi^2 - (\de \zeta - \cosh \vartheta \de  \chi)^2 \right)\,, \\
    \de s^2(S_j^3) &= \frac{1}{4} \left( \de \theta_j^2 + \sin^2 \theta_j \de \eta_j^2 + (\de \xi_j - \cos \theta_j \de  \eta_j)^2 \right)\,, \qquad j=1,2\,, \\
    \de s^2(S^1) &= \de x^2\,.
\end{align}
This geometry features seven manifest $U(1)$ isometries. The coordinates $\chi$ and $\zeta$ realise the $U(1)_L \subset SU(1,1)_L$ and $U(1)_R \subset SU(1,1)_R$ in $AdS_3$. The coordinates $\eta_j \in [0,2\pi)$ correspond to the left $U(1)_{j,L} \subset SU(2)_{j,L}$, while the coordinates $\xi_j \in [0, 2\pi)$ correspond to the right $U(1)_{j,R} \subset SU(2)_{j,R}$
Finally, $x$ is the coordinate on the $S^1$ factor. The other variables have range $\theta_j \in [0,\pi)$. The perhaps more familiar global coordinates of $AdS_3$ and $S^3$ are related to the ones above through 
\begin{equation}
\begin{aligned}
t &=  \frac{\chi-\zeta}{2}~, \qquad &&\psi = \frac{\chi+\zeta}{2}~, \qquad &&\rho =\sinh \frac{\vartheta}{2}\,, \\
    \varphi_j &= \frac{\eta_j-\xi_j}{2}~, \qquad &&\phi_j = \frac{\eta_j+\xi_j}{2}~, \qquad &&r_j =\sin \frac{\theta_j}{2}\,,
    \end{aligned}
\end{equation}
so that
\begin{equation}
\begin{aligned}
\de s^2(AdS_3) &= -(1+\rho^2) \de t^2 + \frac{\de\rho^2}{1+\rho^2} + \rho^2 \de \psi^2\,, \\
    \de s^2(S_j^3) &=+ (1-r_j^2) \de \varphi_j^2 + \frac{\de r_j^2}{1-r_j^2} + r_j^2 \de \phi_j^2\,.
    \end{aligned}
\end{equation}
We are interested in TsT transformations involving  $\xi_1$, $\xi_2$ and $x$, which  leave the $AdS_3$ factor untouched and preserve the left-acting bosonic symmetries. Having three isometries at our disposal we can construct a family of $3(3-1)/2=3$ deformations. Taking into account that we start from a background with two parameters $q$ and $\alpha$, considering its three-parameter deformation becomes a little unwieldy. We will instead proceed as follows:
\begin{itemize}
    \item Consider a TsT in $(\xi_2,x)$, which provides an integrable background with one additional parameter, presented in Section \ref{sec:AB}; as we will see this interpolates between cases 1.\ and 2.;
    \item Consider a TsT in $(\xi_1,\xi_2)$, which also provides an integrable background with one additional parameter, interpolating between 1. and 3., see Section~\ref{sec:AC}.
    \item Consider the most general three-parameter TsT, which we discuss in Section~\ref{sec:ABC}.
\end{itemize}

To write down the deformed metric in the next sections, we find it convenient to introduce the metric of the squashed three-sphere,
\begin{equation}
\label{eq:squashed}
    \de s^2(S_{j,\Delta}^3) = \frac{1}{4} \left( \de s^2(S_j^2)+ \Delta A_j^2 \right)\,,
\end{equation}
with $0<\Delta\leq1$ the deformation parameter, and
\begin{equation}
     \de s^2(S_j^2) = \de\theta_j^2 + \sin^2 \theta_j \de\eta_j^2 \,, \qquad  A_j = \de \xi_j - \cos \theta_j \de \eta_j\,.
\end{equation}
The various volume forms are given by
\begin{align}
\Omega(AdS_3) &= \frac{1}{8} \sinh \vartheta \de\vartheta \wedge \de\chi \wedge \de\zeta = \rho \, \de \rho \wedge \de t \wedge \de \psi\, \\
    \Omega(S_j^3) &= \frac{1}{8} \sin \theta_j \de\theta_j \wedge \de\eta_j \wedge \de\xi_j = r_j \, \de r_j \wedge \de \varphi_j \wedge \de \phi_j\,, \\
    \Omega(S_j^2) &= \sin \theta_j \de\theta_j \wedge \de\eta_j = 4 r_j \de r_j \wedge (\de \varphi_j + \de \phi_j)~,
\end{align}
and we define the B-field
\begin{equation}
\label{eq:Bq}
    B_q = \frac{q}{4} \left[ R^2\cosh\vartheta\, \de \chi \wedge \de \zeta - R_1^2 \cos\theta_1\, \de\eta_1 \wedge \de\xi_1 - R_2^2 \cos \theta_2 \de\eta_2 \wedge \de\xi_2\right]\,,
\end{equation}
so that 
\begin{equation}
    H_q = \de B_q = 2 q \left(R^2\, \Omega(AdS_3) + R_1^2\, \Omega(S_1^3) + R_2^2\,\Omega(S_2^3)\right)\,.
\end{equation}

\section{Integrable interpolation 1.~\texorpdfstring{$\boldsymbol\to$}{->}~2.}
\label{sec:AB}
Here we consider the TsT transformation that interpolates between cases 1.\ and 2.\ of Table~\ref{tab:backgrounds}. We construct the background and study its pp-wave limit when expanding it around a suitable geodesics.

\subsection{Deformed background}
\paragraph{Pure-RR case.}
Let us start from the pure-RR background and consider the following sequence of transformations:
\begin{enumerate}
    \item T-duality in $\xi_2$;
    \item A shift
\begin{equation}
\label{eq:ABshiftRR}
\begin{pmatrix}
    \xi_2 \\ x
    \end{pmatrix} \rightarrow \begin{pmatrix}
    1 & 0 \\
    s & 1 \\
        \end{pmatrix}
    \begin{pmatrix}
    \xi_2 \\ x
    \end{pmatrix}~, \qquad s=\frac{2}{R_2 } \frac{\sqrt{1-\Delta}}{\sqrt{\Delta}}~, \qquad 0 < \Delta \leq 1;
\end{equation}
\item T-duality in $\xi_2$;
\item For convenience, a rescaling of the circle coordinate $x \rightarrow \Delta^{-1/2}x$.
\end{enumerate}
The resulting metric and B-field are
\begin{align}
    \de s^2 &= R^2 \de s^2(AdS_3) + R_1^2 \de s^2(S_1^3) + R_2^2 \de s^2(S_{2,\Delta}^3) + \de x^2~, \\
    B&= \frac{1}{2} R_2 \sqrt{1-\Delta} A_2 \wedge \de x~.
\end{align}
The dilaton and RR fluxes are 
\begin{equation}
\begin{aligned}
    \Phi &= \Phi_0 + \log[\sqrt{\Delta}]~, \qquad F_1 =0~, \\
    F_3 &= e^{-\Phi_0} 2\left(R^2 \Omega(AdS_3) +  R_1^2 \Omega(S_1^3) + R_2^2 \Omega(S_2^3)\right)~, \\
    F_5 &= -e^{-\Phi_0}R_2\sqrt{1-\Delta}(1+\star)  \left( R^2 \Omega(AdS_3) + R_1^2 \Omega(S_1^3) \right) \wedge A_2 \wedge \de x~.
    \end{aligned}
\end{equation}
\paragraph{Strong deformation limit.}
The undeformed background is the one where $s=0$, that is $\Delta=1$. The opposite limit, $s\to+\infty$, corresponds to taking $\Delta\to0$.
In this limit, it is convenient to perform a further rescaling
\begin{equation}
    \xi_2 \rightarrow \frac{2}{\sqrt{\Delta}} \frac{\xi_2}{R_2}~, \qquad \Phi_0 \rightarrow \Phi_0 - \log[\sqrt{\Delta}]~, \qquad \Delta \rightarrow 0~,
\end{equation}
after which the background becomes
\begin{align}
    \de s^2 &= R^2 \de s^2(AdS_3) + R_1^2 \de s^2(S_1^3) + \frac{R_2^2}{4} \de s^2(S_2^2) + \de \xi_2^2 + \de x^2~, \\
    H_3 &= \frac{R_2}{2} \Omega(S_2^2) \wedge \de x~, \\
    F_3 &= e^{-\Phi_0} \frac{R_2}{2} \Omega(S_2^2) \wedge \de \xi_2~, \\
    F_5 &= -e^{-\Phi_0} 2 (1+\star)  \left( R^2 \Omega(AdS_3) + R_1^2 \Omega(S_1^3) \right) \wedge \de\xi_2 \wedge \de x~.
\end{align}
This is precisely the background \ref{eq:background2} of $AdS_3 \times S_1^3 \times S_2^2 \times T^2$.

\paragraph{Mixed-flux case.}
Here we start from a background with a generic $q$, $0\leq q\leq1$, and apply a TsT transformation. The steps 1--3 are the same as above; in particular, the shift is still given by~\eqref{eq:ABshiftRR}, though we replace $\Delta$ with $\Delta'$ for later convenience. This yields
\begin{equation}
\begin{aligned}
    \de s^2 &= R^2 \de s^2(AdS_3) + R_1^2 \de s^2(S_1^3) + R_2^2 \de s^2(S_{2,\Delta'}^3) + \left(\frac{q R_2}{2} \sqrt{1-\Delta'} \cos \theta_2 \de\eta_2 -\sqrt{\Delta'}\de x \right)^2  \,, \\
    B&= B_q +\frac{1}{2} R_2 \sqrt{\Delta'(1-\Delta')} A_2 \wedge \de x + \frac{1}{4} q R_2^2 (1-\Delta') \cos \theta_2 \de \eta_2 \wedge \de \xi_2\,,
\end{aligned}
\end{equation}
where $B_q$ is the B-field before any deformation, given in~\eqref{eq:Bq}.
The metric can be cast in the same form as above
\begin{align}
    \de s^2 &= R^2 \de s^2(AdS_3) + R_1^2 \de s^2(S_1^3) + R_2^2 \de s^2(S_{2,\Delta}^3)+ \de x^2\,,
\end{align}
if, instead of just rescaling $x$, we perform a field redefinition and introduce $\Delta(q,\Delta')$:
\begin{equation}
    \begin{pmatrix}
    \xi_2 \\ x
    \end{pmatrix} \rightarrow \begin{pmatrix}
    1 & -\frac{q s}{\sqrt{\Delta}} \\
    q \frac{R_2^2}{4} s & \frac{1}{\sqrt{\Delta}} \\
        \end{pmatrix}
    \begin{pmatrix}
    \xi_2 \\ x
    \end{pmatrix}\,, \qquad \Delta = \Delta'(1-q^2) + q^2 = \frac{4 + s^2 q^2 R_2^2}{4+s^2 R_2^2}\,.
\end{equation}
Note that now
\begin{equation}
    q^2 \leq \Delta  \leq 1\,,
\end{equation}
so that the maximal deformation $s\to\infty$ corresponds now to $\Delta\to q^2$. In particular, we cannot obtain the $AdS_3 \times S_1^3 \times S_2^2 \times T^2$ geometry starting from a $q>0$ background.
The B-field after field redefinition is
\begin{equation}
\begin{aligned}
    B =& B_q + \frac{R_2 \sqrt{\Delta} \sqrt{1-\Delta}}{2 \sqrt{\Delta-q^2}} \left(\de \xi_2 - \frac{\Delta- q^2}{\Delta} \cos \theta_2 \de \eta_2\right) \wedge \de x,
\end{aligned}
\end{equation}
so that 
\begin{equation}
    H_3=\de B = 2 q \left[ R^2 \Omega(AdS_3) +  R_1^2 \Omega(S_1^3)+ R_2^2 \Omega(S_2^3) \right] + \frac{R_2}{2} \sqrt{\frac{(1-\Delta)(\Delta-q^2)}{\Delta}} \Omega(S_2^2) \wedge \de x.
\end{equation}
The remaining fluxes are
\begin{equation}
    \begin{aligned}
        F_1 &=0\, \\
        F_3 &= \sqrt{1-q^2} e^{-\Phi_0} 2\left[R^2 \Omega(AdS_3) + R_1^2 \Omega(S_1^3) + R_2^2 \Omega(S_2^3) - q
        \frac{R_2}{4}\sqrt{\frac{(1-\Delta)}{\Delta(\Delta-q^2)}}
        \Omega(S_2^2) \wedge \de x\right]\,\\
        F_5 &= - \sqrt{1-q^2} e^{-\Phi_0}  \frac{\sqrt{\Delta} \sqrt{1-\Delta} }{\sqrt{\Delta-q^2}} R_2  (1+\star) (R^2 \Omega(AdS_3) + R_1^2 \Omega(S_1^2) ) \wedge A_2 \wedge \de x\,.
    \end{aligned}
\end{equation}
Note finally that if we start from a pure-NSNS backgound ($q=1$), we again obtain an NSNS background which just differs by swapping the Hopf fiber on $S^{3}_2$ and the circle whose coordinate is~$x$. This changes e.g.\ the string spectrum on this background, but does still yields an $S^2_2$ geometry. 

\subsection{Geodesics}
\label{sec:AB:geodesics}
Let us consider point-like strings, whose motion depends on $\tau$ but not on~$\sigma$, propagating in the above space-time. The geodesic equation for the fields $X^\mu$ is
\begin{equation}
    \ddot{X}^\mu+\Gamma^\mu{}_{\nu\rho}\dot{X}^\nu\dot{X}^\rho=0\,.
\end{equation}
We are interested in solutions of the type
\begin{equation}
    \vartheta=\text{const.}\,,\qquad
    \theta_1=\text{const.}\,,\qquad
    \theta_2=\text{const.}\,, \qquad x = \text{const.}\,.
\end{equation}
Moreover, for the $AdS_3\times S^3$ part of the geometry we will take the usual BMN geodesic, which in Hopf coordinates reads
\begin{equation}
\label{eq:NoSpinAdS3xS3}
    \chi(\tau)=-\zeta(\tau)=c_0\,\tau\,,\quad \vartheta=0\,,\qquad
    \eta_1(\tau)=-\xi_1(\tau)=c_1\,\tau\,,\quad\theta_1=0\,,
\end{equation}
for constants $c_0$ and $c_1$. 
On the remaining part of the geometry the geodesic equations read
\begin{equation}
\label{eq:eom1}
    \ddot{\xi}_2=\ddot{\eta}_2=0\,,\qquad \sin\theta_2\,\dot{\eta}_2\left[(1-\Delta)\cos\theta_2\,\dot{\eta}_2+\Delta\,\dot{\xi}_2\right]=0\,.
\end{equation}
Note that we can describe the deformation of mixed-flux or pure-RR backgrounds simultaneously --- it is just a matter of specifying the range of $\Delta$.  
We solve the first two equations by setting
\begin{equation}
    \eta_2(\tau)=c_2'\,\tau\,,\qquad\xi_2(\tau)=c_2''\,\tau\,,
\end{equation}
while for the remaining equation we have several distinct choices which we will examine later, under the assumption
\begin{equation}
    c_2' > 0\,,
\end{equation}
so that the geodesic runs along~$S^2$.
Finally, the Virasoro constraint is
\begin{equation}
\label{eq:vir1}
    R^2 c_0^2=R_1^2c_1^2+\frac{R_2^2}{4}\left[\sin^2\theta_2\,(c_2')^2+\Delta(c_2''-c_2'\cos\theta_2)^2\right]\,.
\end{equation}
From the Lagrangian of point-like excitations
\begin{equation}
    \mathcal{L}=\frac{1}{2}g_{\mu\nu}\dot{X}^\mu\dot{X}^\nu\,,
\end{equation}
we obtain the conjugate momenta
\begin{equation}
    p_\mu=g_{\mu\nu}\dot{X}^\nu\,,
\end{equation}
and we can therefore relate the quantities introduced earlier to conserved charges. We are interested in particular in the following charges (computed at $\vartheta,\theta_j$ constant)
\begin{equation}
\begin{aligned}
    L_0=-p_\chi&=\frac{R^2}{4}(\dot{\chi}-\dot{\zeta}\cosh\vartheta)\,,\qquad
    &&\bar{L}_0=p_\zeta=\frac{R^2}{4}(\dot{\chi}\cosh\vartheta-\dot{\zeta})\,,\\
    J_1=p_{\eta_1}&=\frac{R_1^2}{4}(\dot{\eta}_1-\dot{\xi}_1\cos\theta_1)\,,\qquad
    &&\bar{J}_1=-p_{\xi_1}=\frac{R_1^2}{4}(\dot{\eta}_2\cos\theta_1-\dot{\xi}_1)\,,\\
    J_2=p_{\eta_2}&=\frac{R_2^2}{4}\left[\dot{\eta}_2\sin^2\theta_2-\Delta \cos\theta_2(\dot{\xi}_2-\dot{\eta}_2\cos\theta_2)\right]\,,\quad
    &&\bar{J}_2=-p_{\xi_2}=\frac{R_2^2}{4}\Delta(\dot{\eta}_2\cos\theta_2-\dot{\xi}_2)\,.
\end{aligned}
\end{equation}
Note that the ansatz~\eqref{eq:NoSpinAdS3xS3} implies
\begin{equation}
    L_0=\bar{L}_0=R^2\frac{c_0}{2}\,,\qquad
    J_1=\bar{J}_1=R_1^2\frac{c_1}{2}\,.
\end{equation}

\paragraph{Constant solution.}
A simple solution to the equation \ref{eq:eom1} is
\begin{equation}
    \theta_2=0\,.
\end{equation}
Let us rewrite
\begin{equation}
    c_0=\frac{2L_0}{R^2}=\frac{2\bar{L}_0}{R^2}\,,\qquad
    c_1=\frac{2J_1}{R^2_1}=\frac{2\bar{J}_1}{R^2_1}\,.
\end{equation}
The remaining two angular momenta are related to the constants $c_2',c_2''$ as
\begin{equation}
    J_2=\bar{J}_2=\frac{R_2^2}{4}\Delta(c_2' -c_2'')\,.
\end{equation}
The Virasoro constraint then reads
\begin{equation}
    \frac{L_0^2}{R^2}=\frac{J_1^2}{R_1^2}+\frac{J_2^2}{\Delta R_2^2}\,,
\end{equation}
which can be parametrised in terms of~$\alpha$ as
\begin{equation}
    L_0^2=\alpha J_1^2+\frac{1-\alpha}{\Delta}\,J_2^2\,,
\end{equation}
and a similar equation holds for the barred quantities.
Recall now that the BPS bound of the $\mathfrak{d}(2,1;\alpha)_L\oplus \mathfrak{d}(2,1;\alpha)_R$ algebra is
\begin{equation}
\label{eq:BPSbound}
    L_0=\alpha\, J_1+(1-\alpha)\,J_2\,,\qquad
    \bar{L}_0=\alpha\, \bar{J}_1+(1-\alpha)\,\bar{J}_2\,.
\end{equation}
Assuming that the BPS bound is satisfied, the Virasoro constraint can be rewritten
\begin{equation}
    0 = (1-\alpha)(1-\Delta) J_2^2 +  \alpha (1-\alpha) \Delta (J_1-J_2)^2\,.
\end{equation}
There is no non-trivial real solution for $0<\Delta<1$ and $0<\alpha<1$.
We conclude that this solution does not preserve any supersymmetry and therefore we discard it.

\paragraph{Two-sphere constant solution.}
If we assume that there is no motion on one of the directions
\begin{equation}
    \dot{\xi}_2=0 \quad \Rightarrow \quad c_2''=0\,,
\end{equation}
then we have a solution with
\begin{equation}
    \theta_2=\frac{\pi}{2}\,.
\end{equation}
The conserved charges are
\begin{equation}
    J_2= \frac{R_2^2}{4}\dot{\eta}_2=\frac{R_2^2}{4}c_2'\,,\qquad \bar{J}_2=0\,,
\end{equation}
so that we set
\begin{equation}
    c_2'=\frac{4 J_2}{R_2^2}\,,\qquad c_2''=0\,.
\end{equation}
The Virasoro constraint becomes
\begin{equation}
    R^2 c_0^2=R_1^2c_1^2+\frac{R_2^2}{4}(c_2')^2\quad\Rightarrow\quad
        L_0^2=\alpha J_1^2+(1-\alpha)\,J_2^2\,,
\end{equation}
which can be solved as usual,
\begin{equation}
    J_1=\bar{J}_1= \sqrt{\frac{\beta}{\alpha}}\,L_0\,,\qquad
    J_2= \sqrt{\frac{1-\beta}{1-\alpha}}\,L_0\,,\qquad 0\leq \beta\leq1\,,\qquad
    \bar{J}_2=0\,.
\end{equation}
In particular, we may choose $\beta=\alpha$ in which case
\begin{equation}
\label{eq:quarterbps}
    J_1=\bar{J}_1=J_2=L_0\,,\qquad \bar{J}_2=0\,.
\end{equation}
The BPS bound is satisfied in the left algebra, but not in the right algebra.
This solution is therefore a candidate for a 1/2-BPS ground state of the left algebra (which is 1/8~BPS in the original underformed geometry).

\paragraph{Continuous solution.}
An interesting family of solutions is
\begin{equation}
c_2''=-\frac{1-\Delta}{\Delta}\,\cos\theta_2\;c_2'\,.
\end{equation}
The angular momenta are
\begin{equation}
    J_2=\frac{R_2^2}{4}\dot{\eta}_2=\frac{R_2^2}{4}c_2'\,,\qquad
    \bar{J}_2=\frac{R_2^2}{4}\cos\theta_2\;\dot{\eta}_2=\frac{R_2^2}{4}\cos\theta_2\;c_2'\,,
\end{equation}
so that
\begin{equation}
\label{eq:JJbarRelation-cont}
    \bar{J}_2=\cos\theta_2\,J_2\,,
\end{equation}
which interpolates between the two cases which we considered above.
We substitute
\begin{equation}
    c_2'=\frac{4 J_2}{R_2^2}\,,
\end{equation}
and the Virasoro constraint reads
\begin{equation}
    L_0^2=\alpha\,J_1^2+(1-\alpha)\,J_2^2\frac{\cos^2\theta_2+\Delta\sin^2\theta_2}{\Delta}.
\end{equation}
The relation for the ``right'' generators follows from~\eqref{eq:JJbarRelation-cont}.
This can be solved by setting
\begin{equation}
    J_1=\bar{J}_1=\sqrt{\frac{\beta}{\alpha}}L_0\,,
\end{equation}
and
\begin{equation}
    J_2=\sqrt{\frac{1-\beta}{1-\alpha}}\sqrt{\frac{\Delta}{\cos^2\theta_2+\Delta\sin^2\theta_2}}L_0\,,\qquad
    \bar{J}_2=\sqrt{\frac{1-\beta}{1-\alpha}}\sqrt{\frac{\cos^2\theta_2\,\Delta}{\cos^2\theta_2+\Delta\sin^2\theta_2}}L_0\,.
\end{equation}
The only solution which allows to solve at least one BPS condition is, for $0<\Delta<1$,
\begin{equation}
    \theta_2=\frac{\pi}{2}\,.
\end{equation}
We conclude that the most convenient geodesic to preserve some supersymmetry of the model is the one considered above, with charges~\eqref{eq:quarterbps}.

\subsection{pp-wave limit}
Informed by the previous discussion we want to take the following geodesic:
\begin{equation}
\dot{\chi}=-\dot{\zeta}=\frac{L_0+\bar{L}_0}{R^2}\,,\qquad
\dot{\eta}_1=-\dot{\xi}_1=\frac{J_1+\bar{J}_1}{R_1^2}\,,\qquad
\dot{\eta}_2=\frac{4J_2}{R_2^2}\,,\qquad
\dot{\xi}_2=0\,,
\end{equation}
subject to the $1/8$-BPS conditions
\begin{equation}
    L_0=\bar{L}_0=J_1=\bar{J}_1=J_2,\qquad\bar{J}_2=0\,.
\end{equation}
This suggests defining the light-cone coordinates
\begin{equation}
\begin{aligned}
    X^+&= +\frac{1}{4}(\chi-\zeta)+\frac{1}{4}(\eta_1-\xi_1)+\frac{1}{4}\eta_2,\\
    X^-&= -\frac{1}{2}(\chi-\zeta)+\frac{1}{2}(\eta_1-\xi_1)+\frac{1}{2}\eta_2,\\
    X^7&= -\frac{1}{2}\sqrt{\frac{1-\alpha}{\alpha}}(\eta_1-\xi_1)+\frac{1}{2}\sqrt{\frac{\alpha}{1-\alpha}}\eta_2,\\
\end{aligned}
\end{equation}
where the factor of~$2$ between $X^+$ and $X^-$ is conventional. For the underformed spaces it is natural to use global coordinates,
\begin{equation}
\begin{aligned}
    X^+&= +\frac{1}{2}t+\frac{1}{2}\varphi_1+\frac{1}{4}\eta_2,\\
    X^-&= -t+\varphi_1+\frac{1}{2}\eta_2,\\
    X^7&= -\sqrt{\frac{1-\alpha}{\alpha}}\varphi_1+\frac{1}{2}\sqrt{\frac{\alpha}{1-\alpha}}\eta_2,\\
\end{aligned}
\end{equation}
and to further perform the change of variables
\begin{equation}
    \rho=\frac{\sqrt{(X^1)^2+(X^2)^2}}{1-\frac{1}{4}[(X^1)^2+(X^2)^2]},\quad \psi=\arctan\frac{X^1}{X^2}\,,\qquad
    r_1=\frac{\sqrt{(X^3)^2+(X^4)^2}}{1+\frac{1}{4}[(X^3)^2+(X^4)^2]}\,,\quad\phi_1=\arctan\frac{X^3}{X^4}\,.
\end{equation}
We also set
\begin{equation}
    \theta_2=\frac{\pi}{2}+\arcsin X^5\,,\qquad\xi_2=X^6\,,
\end{equation}
so that
\begin{equation}
    \frac{R_2^2}{4}\left(\de s^2(S^2)+\Delta A^2\right)= 
    \frac{R_2^2}{4}\left[
    \left(1-(X^5)^2\right) (\de \eta_2)^2+\frac{(\de X^5)^2}{1-(X^5)^2}+\Delta (\de X^6+X^5\de \eta_2)^2
    \right].
\end{equation}
With these conventions we can now perform the pp-wave limit by setting
\begin{equation}
    X^+=x^+\,,\quad X^-=\frac{x^-}{R^2}\,,\quad
    X^{1,2}=\frac{x^{1,2}}{R}\,,\quad
    X^{3,4}=\frac{x^{3,4}}{R_1}\,,\quad
    X^5=\frac{2x^5}{R_2}\,,\quad
    X^6=\frac{2x^6}{\sqrt{\Delta}R_2}\,,
\end{equation}
and taking $R\to\infty$ with $x^\mu$ fixed.
The pp-wave metric reads
\begin{equation}
\de s^2=
2\de x^+\de x^- +2 \kappa\, x^5\de x^6\de x^+ +\delta_{ij}\de x^i\de x^j-A_{ij} x^i x^j(\de x^+)^2\,,
\end{equation}
where
\begin{equation}
    \kappa = 2 (1-\alpha) \sqrt{\Delta}\,, \qquad A_{ij}=\text{diag}(1,1,\alpha^2,\alpha^2,4 (1-\alpha)^2 (1-\Delta),0,0,0)\,.
\end{equation}

The Kalb-Ramond field, before the limit, is
\begin{equation}
\begin{aligned}
    B=&qR^2\left(\frac{1}{2}+\rho^2\right)\de t\wedge\de \psi
    +qR_1^2\left(-\frac{1}{2}+r_1^2\right)\de \varphi_1\wedge\de \phi_1\\
&    + \frac{qR_2^2}{4} X^5 \de\eta_2 \wedge \de X^6
    -\frac{R_2 \sqrt{\Delta} \sqrt{1-\Delta}}{2 \sqrt{\Delta-q^2}} \left(\de X^6 + \frac{\Delta- q^2}{\Delta} X^5 \de \eta_2\right) \wedge \de X^8,
\end{aligned}
\end{equation}
and it is convenient to drop the closed terms before taking the pp-wave limit to avoid generating (closed) divergent terms. Namely, let 
\begin{equation}
\begin{aligned}
    B=&\;qR^2\rho^2\de t\wedge\de \psi
    +qR_1^2r_1^2\de \varphi_1\wedge\de \phi_1\\
&    +\frac{qR_2^2}{4} X^5 \de\eta_2 \wedge \de X^6
    -\frac{R_2}{2}  \frac{\sqrt{1-\Delta}\sqrt{\Delta- q^2}}{\sqrt{\Delta}} X^5 \de \eta_2\wedge \de X^8,
\end{aligned}
\end{equation}
which in the pp-wave limit gives
\begin{equation}
\begin{aligned}
    B=&\;q(x^1\de x^2-x^2\de x^1)\wedge\de x^+
    + \alpha q(x^3\de x^4-x^4\de x^3)\wedge\de x^+ \\
    &-
    \frac{2(1-\alpha)\,q}{\sqrt{\Delta}}x^5\de x^6\wedge \de x^+
    -\frac{2(1-\alpha)\sqrt{1-\Delta}\sqrt{\Delta-q^2}}{\sqrt{\Delta}}x^5\de x^8\wedge\de x^+,
\end{aligned}
\end{equation}
where we recall that $0 \leq q^2 \leq \Delta \leq 1$.

\subsection{Quadratic Hamiltonian}
We are interested in a pp-wave metric of the form
\begin{equation}
G_{--}=0\,,\quad G_{++}=-A_{ij}x^ix^j,\quad
G_{+-}=G_{-+}=1\,,\qquad
G_{+j}=G_{j+}=\delta_{j6} \kappa x^5\,
\qquad G_{ij}=\delta_{ij}\,,
\end{equation}
where the other elements are zero.
The Kalb-Ramond field is of the form
\begin{equation}
B=B_{j+}\,\de x^j\wedge\de x^+\,.
\end{equation}
We first note that the inverse metric is given by
\begin{equation}
G^{++}=0,\quad G^{-+}=G^{+-}=1,\quad G^{--}=A_{ij}x^ix^j+(\kappa x^5)^2,\quad
G^{-6}=G^{6-}=-\kappa x^5\,,\quad G^{ij}=\delta_{ij}\,.
\end{equation}
The form of the quadratic Hamiltonian density
\begin{equation}
\mathcal{H}=-p_{+}\,,
\end{equation}
may be found as usual from the quadratic Virasoro constrain, which in general reads
\begin{equation}
0=G^{\mu\nu}p_{\mu}p_{\nu}+G_{\mu\nu}\acute{x}^\mu\acute{x}^\nu+2G^{\mu\nu}B_{\nu\rho}p_{\mu}\acute{x}^\rho+G^{\mu\nu}B_{\mu\rho}B_{\nu\sigma}\acute{x}^\rho\acute{x}^\sigma\,.
\end{equation}
For a metric of the pp-wave type, $G^{++}=0$ and this can be solved as a linear equation in~$p_+$. Moreover, in the ligthcone gauge
\begin{equation}
x^+=\tau\,,\qquad p_{-}=1\,,
\end{equation}
the last term (quadratic in $B$) also vanishes. We therefore get
\begin{equation}
0=2p_+ +\delta^{ij}p_ip_j+
2G^{-6}p_6 + G^{--}+\delta_{ij}\acute{x}^i\acute{x}^j
+2B_{+\rho}\acute{x}^\rho\,.
\end{equation}
From this we get the Hamiltonian
\begin{equation}
\begin{aligned}
\mathcal{H}=&\;\frac{1}{2}\delta^{ij}p_ip_j+\frac{1}{2}\delta_{ij}\acute{x}^i\acute{x}^j+
\frac{1}{2}\tilde{A}_{ij}x^ix^j-\kappa x^5p_6\\
&-q(x^1\acute{x}^2-x^2\acute{x}^1)-\alpha q(x^3\acute{x}^4-x^3\acute{x}^4)\\
&+\frac{q\kappa}{\Delta}(x^5\acute{x}^6-x^6\acute{x}^5)+
\frac{\kappa\sqrt{1-\Delta}\sqrt{\Delta-q^2}}{\Delta}(x^5\acute{x}^8-x^8\acute{x}^5)\,,
\end{aligned}
\end{equation}
where in the last line we added a total derivative in~$\sigma$, and
\begin{equation}
\begin{aligned}
\tilde{A}_{ij}x^ix^j=&\;{A}_{ij}x^ix^j+\kappa^2(x^5)^2\,.
\end{aligned}
\end{equation}
The diagonalisation of this Hamiltonian gives eight excitations with dispersion relations
\begin{equation}
\begin{aligned}
    &\omega_{1,\pm} = \sqrt{p^2  \pm 2p q+ 1}\,, \qquad \omega_{2,\pm} = \sqrt{p^2  \pm 2\alpha p q + \alpha^2}\,, \\
    &\omega_{3,\pm} = \sqrt{p^2\pm 2 (1-\alpha) pq + (1-\alpha)^2} \pm(1-\alpha) \qquad \omega_{4,\pm} = |p|\,.
    \end{aligned}
\end{equation}
Notice that they do not depend on the deformation parameter $\Delta$. This is actually expected from the fact that  a TsT deformation should result in a twist of the S~matrix, without affecting the quadratic Hamiltonian. Indeed, for the excitations coming from the $AdS_3 \times S_1^3$ geometry we recover the expected dispersion relations $\omega_{1,\pm}$ and $\omega_{2,\pm}$~\cite{Borsato:2015mma}. However, for the excitations coming from the $S_2^3$ geometry, the dispersion relations $\omega_{3,\pm}$ are shifted by a constant, which is precisely the R-charge of that sphere. This shift is due to the fact that we are considering  a different light-cone gauge fixing with respect to the ``usual'' 1/4-BPS geodesics used for $AdS_3\times S^3\times S^3\times S^1$, as discussed in detail in~\cite{Borsato:2023oru} (see also~\cite{Frolov:2019xzi} where these generalised lightcone gauge-fixings were first considered).

\section{Integrable interpolation 1.~\texorpdfstring{$\boldsymbol\to$}{->}~3.}
\label{sec:AC}

In this case we will perform the TsT transformation on the two Hopf fibers of the two three-spheres. The manipulations closely resemble the ones of the previous section, and therefore we will avoid repeating unnecessary details.

\subsection{Deformed background}

\paragraph{Mixed-flux case.}
For brevity, let us right away consider the generic case with $0\leq q\leq 1$; the following formulae immediately yield the deformation of the pure-RR background when setting $q=0$. In analogy with what done before we will perform the following sequence of operations:
\begin{enumerate}
    \item a T-duality in $\xi_1$;
    \item The shift
\begin{equation}
\begin{pmatrix}
    \xi_1 \\ \xi_2
    \end{pmatrix} \rightarrow \begin{pmatrix}
    1 & 0 \\
    s & 1 \\
        \end{pmatrix}
    \begin{pmatrix}
    \xi_1 \\ \xi_2
    \end{pmatrix}~, \qquad s=\frac{4 \sqrt{1-\Delta'}}{R_1 R_2 \sqrt{\Delta'}}~, \qquad  0 < \Delta' \leq 1~;
\end{equation}
\item a T-duality in $\xi_1$.
\end{enumerate}
This gives the metric
\begin{equation}
\begin{aligned}
    \de s^2 &= R^2 \de s^2(AdS_3) + R_1^2 \de s^2(S_{1,\Delta'}^3)+R_2^2 \de s^2(S_{2,\Delta'}^3)+\de x^2 \\
    & + \frac{1}{4} \left( q R_1 \sqrt{1-\Delta'} \cos \theta_1 \de \eta_1 - R_2 \sqrt{\Delta'} \de \xi_2 \right)^2- \frac{R_2^2}{4} \Delta' \de \xi_2^2  \\
    & + \frac{1}{4} \left( q R_2 \sqrt{1-\Delta'} \cos \theta_2 \de \eta_2 + R_1 \sqrt{\Delta'} \de \xi_1 \right)^2 - \frac{R_1^2}{4} \Delta' \de \xi_1^2,
    \end{aligned}
\end{equation}
and B-field
\begin{equation}
\begin{aligned}
    B &= B_q + \frac{1}{4} R_1 R_2 \sqrt{\Delta'(1-\Delta')} A_1 \wedge A_2 \\
    &\quad + \frac{q}{4} R_1^2 (1-\Delta') \cos \theta_1 \de \eta_1 \wedge \de \xi_1 + \frac{q}{4}R_2^2 (1-\Delta') \cos \theta_2 \de \eta_2 \wedge \de \xi_2\\
    &\quad - \frac{q^2}{4} R_1 R_2 \sqrt{\Delta'(1-\Delta')} \cos \theta_1 \cos \theta_2 \de \eta_1 \wedge \de \eta_2,
    \end{aligned}
\end{equation}
where $B_q$ denotes the original B-field given in~\eqref{eq:Bq}.
In the case $q>0$, the metric can be brought to a standard form by means of a further field redefinition
\begin{equation}
    \begin{pmatrix}
    \xi_1 \\ \xi_2
    \end{pmatrix} \rightarrow \begin{pmatrix}
    1 & -\frac{R_2^2}{4}  q s  \\
    \frac{R_1^2}{4} q s & 1 \\
        \end{pmatrix}
    \begin{pmatrix}
    \xi_1 \\ \xi_2
    \end{pmatrix}\,.
\end{equation}
We then obtain
\begin{equation}
\label{eq:ACmetric}
\begin{aligned}
    \de s^2 &= R^2 \de s^2(AdS_3) + R_1^2 \de s^2(S_{1,\Delta}^3)+R_2^2 \de s^2(S_{2,\Delta}^3)+\de x^2\,, \\
    B &= B_q +\frac{1}{4} R_1 R_2 \sqrt{1-\Delta}\sqrt{\Delta-q^2} \left( A_1 \wedge A_2 + \frac{q^2}{\Delta-q^2} \de \xi_1 \wedge \de \xi_2 \right)\,,
    \end{aligned}
\end{equation}
with
\begin{equation}
    \Delta = \Delta' + q^2(1-\Delta')\,, \qquad q^2 < \Delta \leq 1\,.
\end{equation}
The remaining fluxes are
\begin{equation}
   \label{eq:ACRR}
    \begin{aligned}
        F_1 &=0\,, \\
        F_3 &= \sqrt{1-q^2} e^{-\Phi_0} 2\Big[R^2 \Omega(AdS_3) + R_1^2 \Omega(S_1^3) + R_2^2 \Omega(S_2^3)  \\
        &\qquad \qquad +
        q \frac{R_1 R_2}{8} \frac{\sqrt{1-\Delta}}{\sqrt{\Delta-q^2}} \left( \sin \theta_1 \cos \theta_2 \de\theta_1 \wedge \de \eta_1 \wedge \de \eta_2- \cos \theta_1 \sin \theta_2 \de \eta_1 \wedge \de\theta_2 \wedge \de \eta_2 \right)\Big]\,,\\
        F_5 &=  \sqrt{1-q^2} e^{-\Phi_0}  \frac{1}{2}(1+\star) \Omega(AdS_3) \big( -\frac{\sqrt{1-\Delta} }{\sqrt{\Delta-q^2}} \Delta R_1 R_2 A_1 \wedge A_2 \\
        &\qquad \qquad -q \frac{1-\Delta}{1-q^2} R_1^2 \cos \theta_1 \de \eta_1 \wedge \de \xi_1 - q \frac{1-\Delta}{1-q^2} R_2^2 \cos \theta_2 \de \eta_2 \wedge \de \xi_2 \\
        &\qquad \qquad + q^2 R_1 R_2 \frac{1-\Delta}{1-q^2} \frac{\sqrt{1-\Delta}}{\sqrt{\Delta-q^2}}  (\de \xi_1 \wedge \de \xi_2 - \cos \theta_1 \de \eta_1 \wedge \de \xi_2 - \cos \theta_2 \de \xi_1 \wedge \de \eta_2)\big)\,.
    \end{aligned}
\end{equation}

\paragraph{Strong deformation limit.}
Like in the previous section, the limit $s\to\infty$ corresponds to $\Delta\to q^2$. It is only when starting from the pure-RR background, $q=0$, that we can truly interpolate all the way to the $AdS_3\times S^2\times S^2\times T^3$ geometry. In that case we set
\begin{equation}
        \xi_{1,2} \rightarrow \frac{2}{\sqrt{\Delta}} \frac{\xi_{1,2}}{R_{1,2}}~, \qquad \Phi_0 \rightarrow \Phi_0- \log[\sqrt{\Delta}]~, \qquad \Delta \rightarrow 0~,
\end{equation}
so that the background becomes
\begin{align}
        \de s^2 &= R^2 \de s^2(AdS_3) + \frac{R_1^2}{4} \de s^2(S_1^2) + \de\xi_1^2 + \frac{R_2^2}{4} \de s^2(S_2^2)+ \de \xi_2^2 + \de x^2~, \\
        B &=   \frac{\sqrt{1-\Delta}}{\sqrt{\Delta}} \de\xi_1 \wedge \de\xi_2 - \frac{R_2}{2} \cos \theta_2 \de \xi_1 \wedge \de\eta_2 - \frac{R_1}{2} \cos \theta_1 \de \eta_1 \wedge \de\xi_2~.
    \end{align}
    The first contribution in the B-field diverges in the $\Delta \rightarrow 0$ limit, but it is a closed contribution. Indeed,
    \begin{equation}
    \label{eq:ACNSNS}
    \begin{aligned}
        H_3 = \de B = \frac{R_1}{2} \Omega(S_1^2) \wedge \de\xi_2 - \frac{R_2}{2} \Omega(S_2^2) \wedge \de\xi_1~.
        \end{aligned}
    \end{equation}
    For the fluxes we get
    \begin{equation}
        \begin{aligned}
            \Phi&=\Phi_0~, \qquad F_1 =0~,  \\
            F_3 &= e^{-\Phi_0}\left(\frac{R_1}{2} \Omega(S_1^2) \wedge \de\xi_1 + \frac{R_2}{2} \Omega(S_2^2) \wedge \de\xi_2\right)~, \\
            F_5 &= -e^{-\Phi_0} 2 R^2 (1+\star) \Omega(AdS_3) \wedge \de\xi_1 \wedge \de\xi_2~.
        \end{aligned}
    \end{equation}
This is precisely the $AdS_3 \times S^2 \times S^2 \times T^3$ background. Therefore, the metric, B-field \eqref{eq:ACmetric} and RR fluxes \eqref{eq:ACRR} provide an integrable interpolation between the 1.\ and 3. theories, as anticipated.

\subsection{Geodesics}
The study of the geodesics follows closely what sketched in Section~\ref{sec:AB:geodesics}. In this case, we want the usual BMN geodesic on the $AdS_3$ part of the geometry, which in Hopf coordinates reads
\begin{equation}
    \chi(\tau)=-\zeta(\tau)=c_0\,\tau\,,\quad \vartheta=0\,,
\end{equation}
for constant $c_0>0$. 
On the remaining part of the geometry we impose $\theta_j=\text{const.}$, and the geodesic equations read
\begin{equation}
\label{eq:eom2}
    \ddot{\xi}_j=\ddot{\eta}_j=0\,,\qquad \sin\theta_j\,\dot{\eta}_j\left[(1-\Delta)\cos\theta_j\,\dot{\eta}_j+\Delta\,\dot{\xi}_j\right]=0\,,\qquad
    j=1,2\,.
\end{equation}
Again, a linear ansatz is
\begin{equation}
    \eta_j(\tau)=c_j'\,\tau\,,\qquad\xi_2(\tau)=c_j''\,\tau\,,\qquad j=1,2,
\end{equation}
so that the Virasoro constraint is
\begin{equation}
    R^2 c_0^2=\frac{R_1^2}{4}\left[\sin^2\theta_1\,(c_1')^2+\Delta(c_1''-c_1'\cos\theta_1)^2\right]+\frac{R_2^2}{4}\left[\sin^2\theta_2\,(c_2')^2+\Delta(c_2''-c_2'\cos\theta_2)^2\right]\,.
\end{equation}
The conserved charges can also be expressed much like in Section~\ref{sec:AB:geodesics} and they now read
\begin{equation}
\begin{aligned}
    L_0=-p_\chi&=\frac{R^2}{4}(\dot{\chi}-\dot{\zeta}\cosh\vartheta)\,,\,
    &&\bar{L}_0=p_\zeta=\frac{R^2}{4}(\dot{\chi}\cosh\vartheta-\dot{\zeta})\,,\\
    J_j=p_{\eta j}&=\frac{R_j^2}{4}\left[\dot{\eta}_j\sin^2\theta_j+\Delta(\dot{\eta}_j\cos^2\theta_j-\dot{\xi}_j\cos\theta_j)\right]\,,\,
    &&\bar{J}_j=-p_{\xi j}=\frac{R_j^2}{4}\Delta(\dot{\eta}_j\cos\theta_j-\dot{\xi}_j)\,,
\end{aligned}
\end{equation}
where the last line holds for $j=1,2$.

Without repeating the whole analysis of the previous section, it is clear that we can only hope to satisfy the BPS bound~\eqref{eq:BPSbound} in the ``left'' algebra, and this can be done by setting
\begin{equation}
    L_0= \alpha\,J_1+(1-\alpha)J_2\,,\qquad
    \bar{J}_1=\bar{J}_2=0\,,
\end{equation}
that is by setting
\begin{equation}
    c''_j=0\,,\qquad\theta_j=\frac{\pi}{2}\,,\qquad j=1,2\,.
\end{equation}
The Virasoro constraint and the BPS bound then give
\begin{equation}
    c_1'=2\alpha\, c_0\,,\qquad
    c_2'=2(1-\alpha)c_0\,,
\end{equation}
so that
\begin{equation}
\label{eq:chargesAC}
    L_0=J_1=J_2=\bar{L}_0=\frac{c_0}{2}R^2\,,\qquad
    \bar{J}_1=\bar{J}_2=0\,.
\end{equation}

\subsection{pp-wave limit}
As discussed above, the 1/8 BPS geodesic corresponds to
\begin{equation}
\dot{\chi}=-\dot{\zeta}=\frac{L_0+\bar{L}_0}{R^2}\,,\qquad
\dot{\eta}_j=\frac{4J_j}{R_j^2}\,,\qquad
\dot{\xi}_j=0\,,\qquad j=1,2\,,
\end{equation}
subject to the $1/8$-BPS conditions~\eqref{eq:chargesAC}.
This suggests fixing
\begin{equation}
\label{eq:ppAC}
\begin{aligned}
    X^+&= +\frac{1}{2}t+\frac{1}{4}\eta_1+\frac{1}{4}\eta_2,\\
    X^-&= -t+\frac{1}{2}\eta_1+\frac{1}{2}\eta_2,\\
    X^7&= -\frac{1}{2}\sqrt{\frac{1-\alpha}{\alpha}}\eta_1+\frac{1}{2}\sqrt{\frac{\alpha}{1-\alpha}}\eta_2,\\
\end{aligned}
\end{equation}
and to further perform the change of variables
\begin{equation}
\label{eq:ppAC2}
    \rho=\frac{\sqrt{(X^1)^2+(X^2)^2}}{1-\frac{1}{4}[(X^1)^2+(X^2)^2]},\qquad \psi=\arctan\frac{X^1}{X^2}\,,
\end{equation}
and
\begin{equation}
\label{eq:ppAC3}
    \theta_1=\frac{\pi}{2}+\arcsin X^3\,,\qquad\xi_1=X^4\,,\qquad
    \theta_2=\frac{\pi}{2}+\arcsin X^5\,,\qquad\xi_2=X^6\,.
\end{equation}
With these conventions we can now perform the pp-wave limit by setting
\begin{equation}
    X^+=x^+\,,\qquad X^-=\frac{x^-}{R^2}\,,\qquad
    X^{1,2}=\frac{x^{1,2}}{R}\,,
\end{equation}
and for convenience including the rescaling
\begin{equation}
    X^3=\frac{2x^3}{R_1}\,,\qquad
    X^4=\frac{2x^4}{\sqrt{\Delta}R_1}\,,
    \qquad
    X^5=\frac{2x^5}{R_2}\,,\qquad
    X^6=\frac{2x^6}{\sqrt{\Delta}R_2}\,,
\end{equation}
and taking $R\to\infty$ with $x^\mu$ fixed.
The pp-wave metric reads
\begin{equation}
\de s^2=
2\de x^+\de x^- +2 \kappa_1\, x^3\de x^4\de x^++2 \kappa_2\, x^5\de x^6\de x^+ +\delta_{ij}\de x^i\de x^j-A_{ij} x^i x^j(\de x^+)^2\,,
\end{equation}
where
\begin{equation}
    \kappa_1 = 2 \alpha \sqrt{\Delta}\,,\quad
    \kappa_2 = 2 (1-\alpha) \sqrt{\Delta}\,,\qquad A_{ij}=\text{diag}(1,1,4\alpha^2(1-\Delta),0,4 (1-\alpha)^2 (1-\Delta),0,0,0)\,.
\end{equation}
The B-field before the limit is
\begin{equation}
\begin{aligned}
    B=&qR^2\left(\frac{1}{2}+\rho^2\right)\de t\wedge\de \psi
    +\frac{qR_1^2}{4} X^3 \de\eta_1 \wedge \de X^4+\frac{qR_2^2}{4} X^5 \de\eta_2 \wedge \de X^6\\
&    +\frac{1}{4} R_1 R_2 \sqrt{1-\Delta}\sqrt{\Delta-q^2} \left( A_1 \wedge A_2 + \frac{q^2}{\Delta-q^2} \de \xi_1 \wedge \de \xi_2 \right) ,
\end{aligned}
\end{equation}
and it gives
\begin{equation}
\begin{aligned}
B=&\,q (x^1\de x^2-x^2\de x^1)\wedge \de x^+ - \frac{q\alpha}{\sqrt{\Delta}}(x^3\de x^4-x^4\de x^3)\wedge \de x^+ - \frac{q(1-\alpha)}{\sqrt{\Delta}}(x^5\de x^6-x^6\de x^5)\wedge \de x^+\\
&-\frac{\sqrt{1-\Delta}\sqrt{\Delta-q^2}}{\sqrt{\Delta}}\left[
(1-\alpha)(x^4\de x^5-x^5\de x^4)+\alpha (x^3\de x^6-x^6\de x^3)
\right]\wedge\de x^+\,,
\end{aligned}
\end{equation}
where we discarded several closed forms, including  a term proportional to $\de x^4\wedge \de x^6$.

\subsection{Quadratic Hamiltonian}
With respect to the previous case, here the pp-wave metric takes a slightly more general form, as two more off-diagonal elements are  non zero
\begin{equation}
    G_{+j}=G_{j+}=\delta_{j4} \kappa_1 x^3+\delta_{j6} \kappa_2 x^5\,,
\end{equation}
while like before the remaining non-vanishing elements are
\begin{equation}
G_{--}=0\,,\quad G_{++}=-A_{ij}x^ix^j,\quad
G_{+-}=G_{-+}=1\,,
\qquad G_{ij}=\delta_{ij}\,.
\end{equation}
The Kalb-Ramond field is again of the form
\begin{equation}
B= B_{j+}\,\de x^j\wedge\de x^+\,.
\end{equation}
The inverse metric has elements
\begin{equation}
G^{--}=A_{ij}x^ix^j+(\kappa_1 x^3)^2+(\kappa_2 x^5)^2,\qquad
G^{-4}=G^{4-}=-\kappa_1 x^3\,,\qquad
G^{-6}=G^{6-}=-\kappa_2 x^5\,,
\end{equation}
and as usual
\begin{equation}
    G^{++}=0,\qquad G^{-+}=G^{+-}=1\,,\qquad G^{ij}=\delta_{ij}\,.
\end{equation}

The computation of the light-cone gauge-fixed Hamiltonian $\mathcal{H}=-p_{+}$ is similar to what we have already seen and it boils down to solving the linear equation
\begin{equation}
0=2p_+ +\delta^{ij}p_ip_j+
2G^{-4}p_4+
2G^{-6}p_6 + G^{--}+\delta_{ij}\acute{x}^i\acute{x}^j
+2B_{+\rho}\acute{x}^\rho\,.
\end{equation}
From this we get the Hamiltonian
\begin{equation}
\begin{aligned}
\mathcal{H}=&\;\frac{1}{2}\delta^{ij}p_ip_j+\frac{1}{2}\delta_{ij}\acute{x}^i\acute{x}^j+
\frac{1}{2}\tilde{A}_{ij}x^ix^j-\kappa_1 x^3p_4-\kappa_2 x^5p_6\\
&-q(x^1\acute{x}^2-x^2\acute{x}^1)+\frac{q\kappa_1}{2\Delta} (x^3\acute{x}^4-x^4\acute{x}^3)
+\frac{q\kappa_2}{2\Delta} (x^5\acute{x}^6-x^6\acute{x}^5)\\
&+
\frac{\sqrt{1-\Delta}\sqrt{\Delta-q^2}}{2\Delta}\left[
\kappa_1(x^3\acute{x}^6-x^6\acute{x}^3)+
\kappa_2(x^4\acute{x}^5-x^5\acute{x}^4)\right]\,,
\end{aligned}
\end{equation}
where
\begin{equation}
\begin{aligned}
\tilde{A}_{ij}x^ix^j=&\;{A}_{ij}x^ix^j+(\kappa_1x^3)^2+(\kappa_2x^5)^2\,.
\end{aligned}
\end{equation}
The diagonalisation of this Hamiltonian gives excitations with dispersion relations
\begin{equation}
\begin{aligned}
    &\omega_{1,\pm} = \sqrt{p^2  \pm 2p q+ 1}\,, \qquad \omega_{2,\pm} = \sqrt{p^2  \pm 2\alpha p q + \alpha^2} \pm \alpha\,, \\
    &\omega_{3,\pm} = \sqrt{p^2 \pm2 (1-\alpha) pq + (1-\alpha)^2} \pm(1-\alpha)\,, \qquad \omega_{4,\pm} = |p|\,.
    \end{aligned}
\end{equation}
Like before, the only differences with the undeformed dispersion relations are constant shifts by the R-charge of either sphere, due to the different choice of gauge fixing~\cite{Frolov:2019xzi,Borsato:2023oru}.

\section{Three-parameter deformation}
\label{sec:ABC}

In this case we will perform the most general three-parameter deformation that preserves the left supersymmetry.

\subsection{Deformed background}

\paragraph{Pure-RR case.}
For simplicity, and because the formulae are already quite unwieldy in this case, we restrict to the deformation of backgrounds with $q=0$. The three-parameter deformation can be constructed in several ways, which are equivalent up to redefining the deformation parameters. Here we perform the following sequence of transformations
\begin{enumerate}
    \item a T-duality in $\xi_1$;
    \item The shift
\begin{equation}
\begin{pmatrix}
    \xi_1 \\ x
    \end{pmatrix} \rightarrow \begin{pmatrix}
    1 & 0 \\
    s & 1 \\
        \end{pmatrix}
    \begin{pmatrix}
    \xi_1 \\ x
    \end{pmatrix}~, \qquad s=\frac{2\sqrt{1-\Delta_1'}}{R_1 \sqrt{\Delta_1'}}~, \qquad  0 < \Delta_1' \leq 1~;
\end{equation}
\item a T-duality in $\xi_1$.
\item For convenience, a rescaling of the circle coordinate $x \rightarrow (\Delta_1')^{-1/2}x$.
    \item a T-duality in $\xi_2$;
    \item The shift
\begin{equation}
\begin{pmatrix}
    \xi_2 \\ x
    \end{pmatrix} \rightarrow \begin{pmatrix}
    1 & 0 \\
    s & 1 \\
        \end{pmatrix}
    \begin{pmatrix}
    \xi_2 \\ x
    \end{pmatrix}~, \qquad s=\frac{2 \sqrt{1-\Delta_2'}}{R_2 \sqrt{\Delta_2'}}~, \qquad  0 < \Delta_2' \leq 1~;
\end{equation}
\item a T-duality in $\xi_2$.
\item For convenience, a rescaling of the circle coordinate $x \rightarrow (\Delta_2')^{-1/2}x$.
    \item a T-duality in $\xi_1$;
    \item The shift
\begin{equation}
\begin{pmatrix}
    \xi_1 \\ \xi_2
    \end{pmatrix} \rightarrow \begin{pmatrix}
    1 & 0 \\
    s & 1 \\
        \end{pmatrix}
    \begin{pmatrix}
    \xi_1 \\ \xi_2
    \end{pmatrix}~, \qquad s=\frac{4 \sqrt{1-\Delta}}{R_1 R_2 \sqrt{\Delta}}~, \qquad  0 < \Delta \leq 1~;
\end{equation}
\item a T-duality in $\xi_1$.
\end{enumerate}
This gives the metric
\begin{equation}
\label{eq:G3param}
\begin{aligned}
    \de s^2 =\;& R^2 \de s^2(AdS_3) + R_1^2 \de s^2(S_{1,\Delta_1}^3) + R_2^2 \de s^2(S_{2,\Delta_2}^3)\\
    &+ R_1 \beta_1 A_1 \de x - R_2 \beta_2 A_2 \de x - \frac{R_1 R_2}{2} \beta_{12}A_1 A_2 + \frac{1}{D} \de x^2\,,
\end{aligned}
\end{equation}
and B-field
\begin{equation}
\label{eq:B3param}
    B = \frac{1}{2} R_1 b_1 A_1 \wedge \de x +\frac{1}{2} R_2  b_2 A_2 \wedge \de x + \frac{R_1 R_2}{4} b_{12} A_1 \wedge A_2~,
\end{equation}
where we introduced a number of parameters $(\Delta_1,\Delta_2,\beta_1,\beta_2,\beta_{12},b_1,b_2,b_{12},D)$ to simplify the expression, see appendix~\ref{app:3param}. 
In the same appendix, we also collect the formulae for the RR fluxes and illustrate how the metric can be somewhat simplified by redefining the isometric coordinates.

\paragraph{Strong deformation limit.}
From the form of the metric we see that to recover the $S^2$ metric from the squashed sphere one we need to take $\Delta_1\to0$, $\Delta_2\to0$, or both. With three parameters at our disposal, this can be done in a  number of ways. Using the relations of appendix~\ref{app:3param} for instance, we can see that for $\Delta_1'$ and $\Delta_2'$ fixed, it is possible to send both $\Delta_1\to0$ and $\Delta_2\to0$ by taking $\Delta\to0$. It is also possible to take only $\Delta_2\to0$ by taking $\Delta_2'\to0$, whereas to take $\Delta_1\to0$ is is necessary to take $\Delta_1'\to (\Delta_2'-1)/\Delta_2'$ in the three-parameter model (this asymmetry is due to the order in which we performed the TsT transformations). Depending on the specific choice, it may be necessary to rescale the $\mathfrak{u}(1)$ fibers to obtain the $AdS_3\times S^2\times S^2\times T^3$ (or $AdS_3\times S^3\times S^2\times T^2$) metric in the standard form.

\subsection{Geodesics}
Based on the discussion of the two previous subsections, we will forego a detailed discussion of the geodesics but instead focus on constructing a geodesic which has charges
\begin{equation}
    L_0=\bar{L}_0=J_1=J_2\,,\qquad
    \bar{J}_1=\bar{J}_2=0\,,
\end{equation}
precisely like in~\eqref{eq:chargesAC}, so that it satisfies the BPS condition of the left copy of $\mathfrak{d}(2,1;\alpha)$. The Noether charges are given by
\begin{equation}
\begin{aligned}
    L_0=-p_\chi&=\frac{R^2}{4}(\dot{\chi}-\dot{\zeta}\cosh\vartheta)\,,\qquad
    \bar{L}_0=p_\zeta=-\frac{R^2}{4}(\dot{\zeta}-\dot{\chi}\cosh\vartheta)\,,\\
    J_1=p_{\eta_1}&=\frac{R_1^2}{4}\left[\dot{\eta}_1\sin^2\theta_1-\Delta_1 \cos \theta_1 (\dot{\xi}_1-\dot{\eta}_1\cos\theta_1)\right] - \frac{R_1}{2} \beta_1 \cos \theta_1 \dot{x} + \frac{R_1 R_2}{4} \beta_{12} \cos \theta_1 ( \dot{\xi}_2 - \dot{\eta}_2\cos \theta_2 )\,,\\
    \bar{J}_1 = -p_{\xi_1}&=-\frac{R_1^2}{4}\Delta_1(\dot{\xi}_1-\dot{\eta}_1\cos\theta_1) - \frac{R_1}{2} \beta_1 \dot{x}+\frac{R_1 R_2}{4} \beta_{12} ( \dot{\xi}_2 - \dot{\eta}_2\cos \theta_2 ) \,, \\
    J_2=p_{\eta_2}&=\frac{R_2^2}{4}\left[\dot{\eta}_2\sin^2\theta_2-\Delta_2 \cos \theta_2 (\dot{\xi}_2-\dot{\eta}_2\cos\theta_2)\right] + \frac{R_2}{2} \beta_2 \cos \theta_2 \dot{x} + \frac{R_1 R_2}{4} \beta_{12} \cos \theta_2 ( \dot{\xi}_1 - \dot{\eta}_1\cos \theta_1 )\,,\\
    \bar{J}_2 = -p_{\xi_2}&=-\frac{R_2^2}{4}\Delta_2(\dot{\xi}_2-\dot{\eta}_2\cos\theta_2) + \frac{R_2}{2} \beta_2 \dot{x}+ \frac{R_1 R_2}{4} \beta_{12} ( \dot{\xi}_1 - \dot{\eta}_1\cos \theta_1 ) \,, 
\end{aligned}
\end{equation}
and despite the rather cumbersome expression we see that, in the notation of the previous subsection, we can still pick a geodesics with
\begin{equation}
     \dot{\eta}_j=c_j',\qquad \dot{\xi}_j=0\,,\qquad\theta_j=\frac{\pi}{2}\,,\qquad j=1,2\,,
\end{equation}
while the $AdS_3$ part is as usual
\begin{equation}
    \chi(\tau)=-\zeta(\tau)=c_0\,\tau\,,\quad \vartheta=0\,.
\end{equation}
The Virasoro constraint
\begin{equation}
    R^2 c_0^2 = \frac{R_1^2}{4} (c_1')^2 + \frac{R_2^2}{4} (c_2')^2\,,
\end{equation}
and the BPS condition can both be satisfied by picking
\begin{equation}
    c_1'=2\alpha\, c_0\,,\qquad
    c_2'=2(1-\alpha)c_0\,,
\end{equation}
exactly like in Section~\ref{sec:AC}. In other words, we can use the very same geodesic as in the previous section.

\subsection{pp-wave limit}
\label{sec:ABC:pp}
Following Section~\ref{sec:AC} we fix an ansatz for the coordinates compatible with our choice of geodesic. The redefinitions are the very same as in eqs.~(\ref{eq:ppAC}--\ref{eq:ppAC3}).
With these conventions we again perform the pp-wave limit by setting
\begin{equation}
    X^+=x^+\,,\qquad X^-=\frac{x^-}{R^2}\,,\qquad
    X^{1,2}=\frac{x^{1,2}}{R}\,,\qquad
     X^3=\frac{2x^3}{R_1}\,,\qquad
      X^5=\frac{2x^5}{R_2}\,,\qquad
      X^7=\frac{x^7}{R}\,.
\end{equation}
However, in order to obtain a metric which is diagonal in the transverse directions of the string (corresponding to $\de x^j$ with $j=1,\dots,8$), it is necessary to perform a further linear redefinition of $X^4$ and $X^6$ in terms of $x^4$ and $x^6$; the explicit formula is rather cumbersome and can be found in eq.~\eqref{eq:ppwave3param}. Finally, for the circle coordinate we have simply $X^8\to x^8$.
With all redefinitions in place, we take $R\to\infty$ with $x^\mu$ fixed.
The pp-wave metric reads
\begin{equation}
\begin{aligned}
\de s^2=&\ 2\de x^+\de x^- +\delta_{ij}\de x^i\de x^j-A_{ij} x^i x^j(\de x^+)^2\\
&+
2(\kappa_1 x^3-\kappa_2 \frac{\beta_{12}}{\sqrt{\Delta_1\Delta_2}} x^5)\de x^4\de x^+
+2 \kappa_2 \frac{\sqrt{\Delta_1\Delta_2-\beta_{12}^2}}{\sqrt{\Delta_1\Delta_2}}x^5\de x^6\de x^+\,,
\end{aligned}
\end{equation}
where
\begin{equation}
    \kappa_1 = 2 \alpha \sqrt{\Delta_1}\,,\quad
    \kappa_2 = 2 (1-\alpha) \sqrt{\Delta_2}\,,
\end{equation}
and
\begin{equation}
    A_{ij} x^i x^j=(x^1)^2+(x^2)^2+4 \alpha^2 (1-\Delta_1)(x^3)^2 + 4 (1-\alpha)^2 (1-\Delta_2)(x^5)^2 + 8 \alpha(1-\alpha) \beta_{12} x^3 x^5\,.
\end{equation}
For the B-field we have
\begin{equation}
\begin{aligned}
    B &= -2 (\alpha \gamma_1 x^3 + (1-\alpha) \gamma_2 x^5 ) \de x^8 \wedge \de x^+ \\
    &\quad + 2 b_3 \left( \frac{1-\alpha}{\sqrt{\Delta_1}} x^5 \de x^4 - \frac{1}{\sqrt{\Delta_1 \Delta_2 - \beta_{12}^2}} ( \alpha \sqrt{\Delta_1} x^3 -(1-\alpha) \frac{\beta_{12}}{\sqrt{\Delta_1}} x^5 ) \de x^6 \right) \wedge dx^+\,,
    \end{aligned}
\end{equation}
where we discarded several closed forms, and used the short-hands $\gamma_{1}$ and~$\gamma_2$, defined in appendix~\ref{app:3param}.

\subsection{Quadratic Hamiltonian}
The computation of the quadratic Hamiltonian follows closely what explained in Sections~\ref{sec:AB} and~\ref{sec:AC}, and therefore we only report the result:
\begin{equation}
    \begin{aligned}
        \mathcal H &= \frac{1}{2} \delta^{ij} p_i p_j + \frac{1}{2} \delta_{ij} \acute{x}^i \acute{x}^j + \frac{1}{2} \tilde{A}_{ij} x^i x^j \\
        &\quad - ( \kappa_1 x^3 - \frac{\beta_{12}}{\sqrt{\Delta_1}\sqrt{\Delta_2}} \kappa_2 x^5 ) p_4 - \kappa_2 \frac{\sqrt{\Delta_1 \Delta_2 - \beta_{12}^2}}{\sqrt{\Delta_1 \Delta_2}} x^5 p_6 \\
        &\quad + \frac{\kappa_1 b_{12}}{2 \sqrt{\Delta_1 \Delta_2 - \beta_{12}^2}} (x^3 \acute{x}^6 - \acute{x}^3 x^6) + \frac{\kappa_2 b_{12}}{2\sqrt{\Delta_1 \Delta_2}} (x^4 \acute{x}^5 - x^5 \acute{x}^4)  \\
        &\quad + \gamma_1 \alpha (x^3 \acute{x}^8 - x^8 \acute{x}^3)
        + \gamma_2 (1-\alpha) (x^5 \acute{x}^8 - x^5 \acute{x}^3) 
         - \frac{\kappa_2 \beta_{12} b_{12}}{2\sqrt{
        \Delta_1 \Delta_2} \sqrt{\Delta_1 \Delta_2 - \beta_{12}^2}} (x^5 \acute{x}^6 -  x^6 \acute{x}^5)\,,
    \end{aligned}
\end{equation}
with
\begin{equation}
    \tilde{A}_{ij} x^i x^j = (x^1)^2 + (x^2)^2 + 4 \alpha^2 (1-\Delta_1) (x^3)^2 + 4 (1-\alpha)^2 (1-\Delta_2) (x^5)^2\,.
\end{equation}
Based on our previous discussion, we expect the dispersion relation not to feature the deformation parameters, and indeed to coincide with the one of $AdS_3\times S^3\times S^3\times S^1$ up to constant shifts due to the different choice of lightcone geodesics. This constitutes a check of our rather involved expressions, and in fact we find
\begin{equation}
\label{eq:omega3param}
\begin{aligned}
    &\omega_{1,\pm} = \sqrt{p^2 + 1}\,, \qquad \omega_{2,\pm} = \sqrt{p^2 +\alpha^2} \pm \alpha\,,\\
    &\omega_{3,\pm} = \sqrt{p^2 + (1-\alpha)^2} \pm (1-\alpha)\,, \qquad \omega_{4, \pm} =|p|\,,
    \end{aligned}
\end{equation}
as it should be.

\section{Relation to trigonometric deformations}
\label{sec:trigonometric}
It is interesting to observe that the metric obtained by the above TsT transformations is closely related to the one of the so-called trigonometric deformation, see e.g.~\cite{Hoare:2023zti}. Indeed, on a single sphere the trigonometric deformation results in a squashed metric like~\eqref{eq:squashed}. In the three-parameter deformation above, we can get a metric with two squashed spheres, with squashing parameters $\Delta$, by setting $\Delta_1=\Delta_2=\Delta$ in eq.~\eqref{eq:G3param}; this also imposes $\beta_1=\beta_2=\beta_{12}=0$.
The key difference between the backgrounds resulting from the above TsT transformations and the ones arising from trigonometric deformations is in the fluxes. In particular, the B-field of the trigonometric deformation \textit{vanishes}, unlike here.
It is interesting to work out the pp-wave Hamiltonian for the trigonometric deformation, and compare it with eq.~\eqref{eq:omega3param}. This can be straightforwardly computed by putting the B-field to zero in the formulae above, yielding
\begin{equation}
\begin{aligned}
&\omega_{1,\pm} = \sqrt{p^2+1}\,, \qquad \omega_{4,\pm} = |p|\,,\\
&\sqrt{\omega_{2,\pm}^2 + \hat{\gamma}_2^2} = \sqrt{p^2 +\hat{\gamma}_1^2} \pm \hat{\gamma}_3\,, \qquad \hat{\gamma}_1 = \frac{\alpha}{\sqrt{\Delta}}\,, \qquad \hat{\gamma}_2 = \frac{\alpha(1-\Delta)}{\sqrt{\Delta}}\,, \qquad \hat{\gamma}_3 = \frac{\alpha}{\sqrt{\Delta}}\,, \\
&\sqrt{\omega_{3,\pm}^2 + \check{\gamma}_2^2} = \sqrt{p^2 +\check{\gamma}_1^2} \pm \check{\gamma}_3\,, \qquad \check{\gamma}_1 = \frac{1-\alpha}{\sqrt{\Delta}}\,, \qquad \check{\gamma}_2 = \frac{(1-\alpha)(1-\Delta)}{\sqrt{\Delta}}\,, \qquad \check{\gamma}_3 = \frac{1-\alpha}{\sqrt{\Delta}}\,.
\end{aligned}
\end{equation}
These results slightly generalise those found in~\cite{Hoare:2023zti} for the trigonometric deformation of a single $S^3$; the latter can be recovered by setting $\alpha\to0$ or $\alpha\to1$. Note that deformation parameters now appear explicitly in the dispersion relation. This is expected, because contrary to the TsT transformation, the trigonometric deformation modifies the geometry locally and is believed to  promote part of the isometries --- in this case, $\mathfrak{su}(2)_{1,R}\oplus \mathfrak{su}(2)_{2,R}$ --- to a quantum group.
We expect that to obtain a consistent supergravity background it will be necessary to deform $\mathfrak{d}(2,1;\alpha)_R$ as a whole, in analogy with what done in~\cite{Hoare:2023zti} (where a whole $\mathfrak{psu}(1,1|2)$ was deformed to obtain a supergravity background), at the price of spoiling the $AdS_3$ piece of the geometry.

\section{Conclusions and outlook}
\label{sec:conclusions}
We have shown that it is possible to construct integrable backgrounds that interpolate between the $AdS_3\times S^3\times S^3\times S^1$ background and either $AdS_3\times S^3\times S^2\times T^2$ or $AdS_3\times S^2\times S^2\times T^3$, up to the global topology of~$T^n$. Our formulae are valid for any $\alpha$, so that we can take the limit where either $S^3$ in $AdS_3\times S^3\times S^3\times S^1$ blows up and immediately obtain the interpolation between 
$AdS_3\times S^3\times T^4$ and
$AdS_3\times S^2\times T^5$.
Interestingly, this procedure works only if we perform the TsT transformation on the pure-RR background. In the mixed-flux case, the fiber in $S^3$ does not decompactify as we take a strong-deformation limit; instead, its radius is of order $R_j/q$ where $R_j$ is the original radius and $0\leq q\leq 1$ labels the mixed-flux backgrounds ($q=0$ for pure-RR, and $q=1$ for pure-NSNS). In particular, for a pure-NSNS background, the TsT cannot be interpreted as a change of the metric like in the RR case. This does not mean that the transformation is trivial, as it will have nonetheless a (simple) effect on the spectrum, which can be computed in the integrability picture by means of a Drinfel'd-Reshetikhin twist.

Of course it would also be possible to consider more general deformations. To begin with, one could deform both copies (left and right) of the supersymmetry algebra, along the lines of~\cite{Hoare:2014oua,Seibold:2019dvf}. Alternatively, one could consider trigonometric or elliptic deformations of any of the $\mathfrak{su}(2)$ algebras (as well as of the $\mathfrak{su}(1,1)$ algebra) appearing in the background. Such deformations were recently considered in~\cite{Hoare:2023zti}. The trigonometric deformation depends on a single parameter and it yields the same squashed metric encountered in this paper, though with a different --- vanishing --- B-field; the elliptic one depends on two parameters. It is believed that the integrable structure of these backgrounds should feature trigonometric and elliptic quantum groups, respectively. They too can be investigated in the richer context of $AdS_3\times S^3\times S^3\times S^1$.
As discussed in Section~\ref{sec:trigonometric}, we expect that to obtain a consistent supergravity background it would be necessary to consider a deformation of the whole $\mathfrak{d}(2,1;\alpha)_R$ algebra, rather than of each or some of its bosonic $\mathfrak{su}(2)$ subalgebras (in analogy~\cite{Hoare:2023zti}).

In this paper we have not carried out a fully-fledged analysis of the quantum integrable model arising from these deformations in the lightcone gauge, but our investigation of the pp-wave limit suggests that, as expected on general grounds~\cite{Frolov:2005dj,Alday:2005ww}, it should be possible to obtain the spectrum of these model by suitably twisting their boundary conditions. It should be possible to implement such twists in the integrability construction for the spectrum along the lines of~\cite{vanTongeren:2021jhh}. The integrability construction of the spectrum was recently worked out for pure-RR $AdS_3\times S^3\times T^4$~\cite{Ekhammar:2021pys,Cavaglia:2021eqr,Frolov:2021bwp,Brollo:2023pkl}, but it is still outstanding for the more general case of $AdS_3\times S^3\times S^3\times S^1$.%
\footnote{The main stumbling block is constructing the dressing factors for the model which may be possible following and generalising the results of~\cite{Frolov:2021fmj}.}
In any case, perhaps starting from the case of the interpolation from $AdS_3\times S^3\times T^4$ to $AdS_3\times S^2\times T^5$, it would be very interesting to get more insight on the integrability structure and perhaps on the dual CFT, see~\cite{Witten:2024yod}. Unfortunately, since we are dealing with backgrounds supported by RR flux, it is going to be very challenging to describe the holographic dual, as it is not going to be given by a free orbifold CFT, see~\cite{Brollo:2023pkl}. It may be interesting to consider the brane construction for these backgrounds, following the ideas of~\cite{Orlando:2010ay} and extending them to the D1-D5-D5' system which yields the $AdS_3\times S^3\times S^3\times S^1$ background as near-horizon limit.

Finally, the interpolating backgrounds constructed here may provide an independent way to solve string theory on the integrable $AdS_3\times S^2\times \mathcal{M}$ backgrounds constructed in~\cite{Wulff:2017zbl,Wulff:2017vhv} --- a class of backgrounds so far rather mysterious --- by building on the growing amount of results for the $AdS_3\times S^3\times \mathcal{M}$ cases. We hope to return to some of these questions in the future.

\section*{Acknowledgments}
We thank the participants of the Workshop \textit{Integrability in Low Supersymmetry Theories} in Trani (Italy) in 2024  for stimulating discussions that initiated and furthered this project. We are also grateful to the Kavli Institute for Theoretical Physics in Santa Barbara for the hospitality during the follow-on of the Integrable22 workshop, where part of this review was written.
FS is supported by the Deutsche Forschungsgemeinschaft (DFG, German Research Foundation) under the Collaborative Research Center 1624 ``Higher structures, moduli spaces and integrability'', project number 506632645. This work has been partially funded by the DFG -- Projektnummer 491245950.
A.S.~acknowledges support from the EU -- NextGenerationEU, program STARS@UNIPD, under project ``Exact-Holography'', and from the PRIN Project n.~2022ABPBEY. A.S.~also acknowledges support from the CARIPLO Foundation under grant n.~2022-1886, and from the CARIPARO Foundation Grant under grant n.~68079.

\appendix

\section{Explicit formulae for the three-parameter deformation}
\label{app:3param}
The relation between the parameters in the metric \ref{eq:G3param}, B-field \ref{eq:B3param} and the parameters involved in the TsT transformations is given by
\begin{equation}
\begin{aligned}
    &D= \Delta + (1-\Delta)\Delta'_1 \Delta'_2~, \qquad \Delta_1 = \frac{\Delta (1-\Delta'_2 + \Delta'_1 \Delta'_2)}{D}~, \qquad \Delta_2 = \frac{\Delta \Delta'_2}{D}~, \\
    &\beta_1 =  \frac{\sqrt{(1-\Delta)\Delta(1-\Delta'_2)}}{D}~, \qquad \beta_2 =  \frac{\sqrt{(1-\Delta)\Delta(1-\Delta'_1)\Delta'_2}}{D}~, \qquad \beta_{12} =  \frac{\Delta\sqrt{(1-\Delta'_1) (1-\Delta'_2)\Delta'_2}}{D}\,, \\
    &b_1 = \frac{\Delta}{D} \sqrt{(1-\Delta_1')\Delta_2'}\,, \qquad b_2 = \frac{\Delta}{D} \sqrt{1-\Delta_2'}\,, \qquad b_{12} = \frac{\sqrt{\Delta(1-\Delta)}}{D}\Delta_1' \Delta_2'\,.
    \end{aligned}
\end{equation}
Note the identities
\begin{equation}
    \Delta_1 \Delta_2 - \beta_{12}^2 = \frac{\Delta^2}{D^2} \Delta'_1 \Delta'_2 \geq 0\,, \qquad \beta_1 \beta_2 = \frac{1-\Delta}{D} \beta_{12}\,, \qquad \Delta_1 + \Delta_2 = \frac{\Delta}{D} (1+\Delta'_1 \Delta'_2)\,.
\end{equation}
Another useful set of parameters is $(\gamma_1, \gamma_2, \Delta)$, related to the shifts in the TsT through
\begin{equation}
    \Delta_2' = 1-\gamma_2^2\,, \qquad \Delta_1'  = 1-\frac{\gamma_1^2}{1-\gamma_2^2}\,, \qquad 0 \leq \gamma_1 \leq \sqrt{1-\gamma_2^2}\, \qquad  0 \leq \gamma_2 \leq 1\,.
\end{equation}
In terms of those, the coefficients in the metric and B-field can be written
\begin{equation}
\begin{aligned}
    &\beta_1 = \gamma_2  \frac{\Delta}{D} \sqrt{\frac{1-\Delta}{\Delta}}\,, \qquad \beta_2 = \gamma_1 \frac{\Delta}{D} \sqrt{\frac{1-\Delta}{\Delta}}\,, \qquad \beta_{12} = \frac{\Delta}{D} \gamma_1 \gamma_2\,, \\
    &b_1 =  \frac{\Delta}{D} \gamma_1\,, \qquad b_2 =  \frac{\Delta}{D} \gamma_2\,, \qquad b_{12} =  \frac{\Delta}{D} \sqrt{\frac{1-\Delta}{\Delta}} (1-\gamma_1^2-\gamma_2^2)\,, \\
    &\Delta_1 = \frac{\Delta}{D} (1-\gamma_1^2)\,, \qquad \Delta_2 = \frac{\Delta}{D} (1-\gamma_2^2)\,, \qquad D = 1-(1-\Delta)(\gamma_1^2+\gamma_2^2) \geq \Delta\,.
    \end{aligned}
\end{equation}
These parameters are particularly convenient to write the fluxes,
\begin{align}
    F_1 &= 0\,, \\
    F_3 &= 2 e^{-\Phi_0} \left( R^2 \Omega(AdS_3) + R_1^2 \Omega(S_1^3) + R_2^2 \Omega(S_2^2) \right)\,, \\
    F_5 &= (1+\star)\big( v_1 R^2 \Omega(AdS_3) \wedge A_1 \wedge dx+  v_2 R^2 \Omega(AdS_3) \wedge A_2 \wedge dx\\
    &\quad - \frac{\Delta R_1}{D} \gamma_1 (R^2 \Omega(AdS_3) + R_2^2  \Omega(S_2^3) ) \wedge A_1 \wedge dx \\
    &\quad - \frac{\Delta R_2}{D} \gamma_2  ( R^2 \Omega(AdS_3) + R_1^2 \Omega(S_1^3)) \wedge A_2 \wedge dx \\
    &\quad - \frac{R_1 R_2}{2} \sqrt{\Delta(1-\Delta)} \frac{1-\gamma_1^2-\gamma_2^2}{D} R^2 \Omega(AdS_3) \wedge A_1 \wedge A_2 \wedge dx \big)\,.
\end{align}
The special cases discussed in Section~\ref{sec:AB} and \ref{sec:AC} correspond to 
\begin{align}
    \Delta=\Delta_2'=1 \qquad \Rightarrow \qquad \gamma_1=\gamma_2=0\,, \qquad \Delta_1 =1\,, \qquad \Delta_2 = \Delta_2'\,, \\
    \Delta_1'=\Delta_2'=1 \qquad \Rightarrow \qquad \gamma_1=\gamma_2=0\,, \qquad \Delta_1 =\Delta\,, \qquad \Delta_2 = \Delta\,.
\end{align}
The terms $A_1 \de x$ and $A_2 \de x$ in the metric can be removed by the field redefinition
\begin{equation}
    (\xi_1, \xi_2) \rightarrow (\xi_1 - \frac{2 \gamma_2}{R_1} \sqrt{\frac{1-\Delta}{\Delta}}  x, \xi_2 + \frac{2 \gamma_1}{R_2} \sqrt{\frac{1-\Delta}{\Delta}}  x)\,,
\end{equation}
so that the metric of the three-parameter deformation simply becomes
\begin{equation}
    ds^2 = R^2 ds^2(AdS_3) + R_1^2 ds^2(S_{1,\Delta_1}^3) + R_2^2 ds^2(S_{2,\Delta_2}^3) - \frac{R_1 R_2}{2} \beta_{12} A_1 A_2 +  dx^2\,.
\end{equation}
Note that this transformation normalises the $dx^2$ term.

Finally, we report here the change of variables needed to bring the metric in the pp-wave form in Section~\ref{sec:ABC:pp} 
\begin{equation}
\label{eq:ppwave3param}
\begin{aligned}
    X^4 &=\frac{2}{R_1}\left(\frac{x^4+\frac{1}{\sqrt{\Delta_1 \Delta_2 - \beta_{12}^2}} x^6}{\sqrt{\Delta_1}}
    +\frac{(\beta_{12}\beta_2-\beta_1\Delta_2)x^8}{\Delta_1\Delta_2-\beta^2}\right)\,, \\
    X^6 &=\frac{2}{R_2}\left(\frac{\sqrt{\Delta_1 \Delta_2}}{\sqrt{\Delta_1 \Delta_2 - \beta_{12}^2}} \frac{x^6}{\sqrt{\Delta_2}}-\frac{(\beta_{12}\beta_2-\beta_2\Delta_1)x^8}{\Delta_1\Delta_2-\beta^2}\right)\,.
\end{aligned}
\end{equation}

\bibliographystyle{JHEP}
\bibliography{refs}

\end{document}